\documentclass{jnmp}

\begin{document}
\setcounter{page}{256}
\newtheorem{lem}{Lemma}[section]
\newtheorem{theor}[lem]{Theorem}
\newtheorem{prop}[lem]{Proposition}

\renewcommand{\evenhead}{S N M Ruijsenaars}
\renewcommand{\oddhead}{Reflectionless Analytic Difference Operators
 II}

\thispagestyle{empty}


\FirstPageHead{8}{2}{2001}
{\pageref{ruijsenaarsII-firstpage}--\pageref{ruijsenaarsII-lastpage}}{Article}

\copyrightnote{2001}{S N M Ruijsenaars}

\Name{Reflectionless Analytic Difference Operators II. \\
Relations to Soliton Systems}\label{ruijsenaarsII-firstpage}

\Author{S N M RUIJSENAARS}

\Address{Centre for Mathematics and Computer Science\\
P.O. Box 94079, 1090 GB Amsterdam, The Netherlands}

\Date{Received July 30, 2000; Accepted January 10, 2001}

\begin{abstract}
\noindent
This is the second part of a series of papers dealing
with an extensive class of analytic difference operators
admitting reflectionless eigenfunctions. In the first
part, the pertinent difference operators and their
reflectionless eigenfunctions are constructed from given
``spectral data'', in analogy with the IST for
reflectionless Schr\"odinger and Jacobi operators. In the
present paper, we introduce a suitable time dependence in
the data, arriving at explicit solutions to a nonlocal
evolution equation of Toda type, which may be viewed as
an analog of the KdV and Toda lattice equations for the
latter operators. As a corollary, we reobtain various
known results concerning reflectionless Schr\"odinger and
Jacobi operators. Exploiting a reparametrization in terms
of relativistic Calogero--Moser systems, we also present a
detailed study of $N$-soliton solutions to our nonlocal
evolution equation.
\end{abstract}

\renewcommand{\theequation}{\thesection.\arabic{equation}}
\setcounter{equation}{0}

\section{Introduction}

In our previous paper Ref.~\cite{ruijsenaars:r=0I} we have introduced
and studied an extensive class of A$\Delta$Os (analytic
difference operators) admitting reflectionless
eigenfunctions. Our construction in Ref.~\cite{ruijsenaars:r=0I} is
patterned after the IST for Schr\"odinger and Jacobi
operators. In brief, we start from given spectral data
$(r,\mu)$, and associate to these data a meromorphic
reflectionless wave function ${\mathcal W}(x,p)$ and meromorphic
coefficients $V_a(x)$, $V_b(x)$ (``potentials'') of an A$\Delta$O
$A$ of the form
\begin{equation}\label{ruijsenaars:A}
A\equiv T_i+V_a(x)T_{-i}+V_b(x),\qquad T_{\pm i}\equiv
\exp (\mp i\partial_x),
\end{equation}
such that we have the eigenvalue equation
\begin{equation}\label{ruijsenaars:Wade}
(A{\mathcal W})(x,p)=\left(e^p+e^{-p}\right){\mathcal W}(x,p).
\end{equation}

The data $r=(r_1,\ldots,r_N)$ allowed in Ref.~\cite{ruijsenaars:r=0I}
(from now on denoted by Part~I) gives rise to the poles
of the transmission coefficient, just as for
Schr\"odinger and Jacobi operators. It consists of
complex numbers restricted by
\begin{equation}\label{ruijsenaars:req1}
\mbox{Im}\; r_n\in (-\pi,0)\cup (0,\pi),\qquad n=1,\ldots, N,
\end{equation}
and
\begin{equation}\label{ruijsenaars:req2}
e^{r_m}\ne e^{\pm r_n},\qquad 1\le m<n\le N.
\end{equation}
This guarantees that the
Cauchy matrix
\begin{equation}\label{ruijsenaars:Cr}
C(r)_{mn}\equiv \frac{1}{e^{r_m}-e^{-r_n}},\qquad
m,n=1,\ldots,N,
\end{equation}
is well defined and regular, just as the Cauchy matrix
pertinent to the Schr\"odinger and Jacobi cases.
(Cf.~Eqs.~(2.30)--(2.33) in Part~I, or, briefly,
I(2.30)--(2.33).)

The ``normalization coefficients'' $\mu
=(\mu_1,\ldots,\mu_N)$ permitted in Part~I are however
far more general than for reflectionless Schr\"odinger and
Jacobi operators. Indeed, they are allowed to be
meromorphic functions satisfying (cf.~I(2.34))
\begin{equation}\label{ruijsenaars:mu}
\mu_n(x+i)=\mu(x),\qquad \lim_{|{\rm Re}\; x|\to \infty} \mu_n(x)=c_n,\qquad
c_n\in{\mathbb C}^*,\qquad n=1,\ldots,N.
\end{equation}

Of course, this includes the constant case
$\mu_n(x)=c_n$, which is the analog of the Schr\"odinger
and Jacobi settings. In Part~III of this series of
papers~\cite{ruijsenaars:r=0III} (which deals with various
functional-analytic features), we focus attention on the
constant multiplier case. In Sections~2 and 5 of the
present paper, however, we allow the same multipliers as
in Part~I.

In Section~2 we introduce time dependence in the
multipliers $\mu_1(x),\ldots,\mu_N(x)$. Accordingly, the
potentials $V_a(x)$, $V_b(x)$ and wave function ${\mathcal W}(x,p)$
depend on time. The time dependence is chosen such that
for each set of ``initial'' spectral data $(r,\mu)$, we
arrive at a solution to a nonlocal Toda type equation.
Specifically, this equation reads
\begin{equation}\label{ruijsenaars:nonto}
\ddot{\Psi}(x,t)=i\exp
(i[\Psi(x+i,t)-\Psi(x,t)])-i\exp(i[\Psi(x,t)-\Psi(x-i,t)]).
\end{equation}
Here, the function $\exp[i\Psi(x,t)]$, $x,t\in {\mathbb C}$, is
assumed to be meromorphic in $x$ for all $t$.
With suitable restrictions on the data $r$, $\mu$, we show
that our solutions to (\ref{ruijsenaars:nonto}) are real-valued,
real-analytic functions for real $x$, $t$, with a solitonic
long-time behavior (in Section~6).

Soliton equations of this nonlocal type have been
introduced and studied before, cf.~Santini's
review Ref.~\cite{ruijsenaars:sant} and references given there. (In
particular, when one replaces the lhs of (\ref{ruijsenaars:nonto}) by
$i\partial_x \dot{\Psi}(x,t)$, then one obtains the
so-called intermediate Toda lattice \cite{ruijsenaars:dega}.)
Even so, it seems that (\ref{ruijsenaars:nonto}) is a new soliton
equation, in the sense that there appears to be no obvious
transformation relating it to previously known soliton
equations.

We should stress at this point that we do not associate
to (\ref{ruijsenaars:nonto}) a clear-cut auxiliary linear spectral
problem for self-adjoint A$\Delta$Os on $L^2({\mathbb R},dx)$. In
fact, we view it as a challenging open problem to do so,
in such a way that the self-adjoint A$\Delta$Os of
Part~III~\cite{ruijsenaars:r=0III} arise as the ones with vanishing
reflection.

In Sections~3 and 4 we present further results that render
the existence of such a scenario plausible. Indeed,
in these sections we show
that our (constant multiplier) A$\Delta$Os can be related to
the well-known reflectionless Schr\"odinger and Jacobi
operators, respectively. Moreover, in Section~4 we show
that the corresponding solutions to (\ref{ruijsenaars:nonto}) obtained
in Section~2 are related in the same way to the
Toda lattice solitons.

Just as Part~I~\cite{ruijsenaars:r=0I}, Sections~2--4 are basically
self-contained, in the sense that we need not invoke any
substantial previous results in the literature.
(To be sure, the previous,~``IST-based'', literature in
the KdV and Toda lattice settings did provide
considerable inspiration for this paper as well as
Part~I.) For example, we obtain the solution property for
the  Toda lattice solitons as a corollary of the
corresponding result in Section~2.

In contrast, Section~5 (and Section~6, too) involves
previous work of ours: We demonstrate there that the
reflectionless wave functions and A$\Delta$Os can be
connected to a special class of relativistic
Calogero--Moser type systems. We have labeled the
pertinent finite-dimensional soliton systems by
$\tilde{{\rm II}}_{{\rm rel}}(\tau =\pi/2)$ in our
paper~\cite{ruijsenaars:aa2}; more general integrable $N$-particle
systems of Calogero--Moser and Toda type are surveyed in
our lecture notes Ref.~\cite{ruijsenaars:scmt}.

This relation plays an important role in Part~III. It
enables us to invoke various results from Ref.~\cite{ruijsenaars:aa2}
to control analytical difficulties. Staying within the
context of the present paper, it yields a map from an
arbitrary point in the $\tilde{{\rm II}}_{{\rm rel}}(\tau
=\pi/2)$ $N$-body phase space to a (real-valued)
$N$-soliton solution of (\ref{ruijsenaars:nonto}). As such, it gives
rise to one more example of what we have dubbed
soliton-particle correspondence in our survey
Ref.~\cite{ruijsenaars:fdss}. In this connection we also mention
previous results on this correspondence for the KdV and
Toda lattice solitons~\cite{ruijsenaars:rusc}, and more recent
results on the relation between relativistic
Calogero--Moser systems and the 2D Toda field
theory~\cite{ruijsenaars:krza,ruijsenaars:kpto}.

A novel feature of the correspondence in the present
setting is that our general $N$-soliton solutions to
(\ref{ruijsenaars:nonto}) are encoded via the {\em defining} Lax
matrix of the $\tilde{{\rm II}}_{{\rm rel}}(\tau
=\pi/2)$ system, and not via the dual (``action-angle'') Lax
matrix (as is the case for the sine-Gordon and modified
KdV particle-like solutions \cite{ruijsenaars:aa2}). To be more
specific, the general $N$-soliton solution consists of
$N_{+}\in \{ 0,1,\ldots,N\}$ solitons moving to the
right, and $N_{-}=N-N_{+}$ solitons moving to the left.
The right-movers and left-movers are parametrized by the
particle and antiparticle variables, respectively. When
$N_{+}$ or $N_{-}$ vanishes, one is dealing with a
self-dual pair of Lax matrices, so this new feature plays
no role.

For $N_{+}N_{-}>0$, however, self-duality breaks
down. In  that case the correspondence between
right-/left-movers and particles/antiparticles is quite
different from the correspondence between solitons,
antisolitons and breathers in the sine-Gordon and
modified KdV settings on the one hand, and particles,
antiparticles, and their bound states in the
$\tilde{{\rm II}}_{{\rm rel}}(\tau
=\pi/2)$ system on the other hand.

The consequences of this novel type of correspondence
are made explicit in Section~6, where we study the
$N$-soliton solutions to (\ref{ruijsenaars:nonto}). First of all, we
demonstrate that the pertinent solutions deserve their
name. Indeed, we show that for long times they can be
approximated by linear combinations of $N$ 1-soliton
solutions; moreover, the asymptotic velocities are
conserved and the position shifts are factorized in
terms of pair shifts.

We obtain the features just mentioned in a quite direct
and elementary fashion, yielding however a weaker
approximation result than we obtained for the
sine-Gordon and (m)KdV solitons in Section~7 of
Ref.~\cite{ruijsenaars:aa2}.  Just as in all
cases studied previously, the long-time behavior of the
$N$-soliton solutions is intimately related to the
spectral asymptotics for $t\to \pm \infty$ of
time-dependent matrices. When $N_{-}$ or $N_{+}$
vanishes, we show that our results in {\it
loc.~cit.}~give rise to a natural notion of global,
non-intersecting, space-time trajectories for the $N$
right-moving or $N$ left-moving solitons.

For
$N_{+}N_{-}>0$, however, we wind up with non-intersecting
space-time trajectories only for sufficiently large
times. In between, trajectories are ill-defined, since a
collision of two trajectories typically gives rise to a
complex-conjugate pair of eigenvalues. From a physical
viewpoint, the right-moving soliton (``particle'') and
left-moving soliton (``antiparticle'') involved in the
collision form a resonance for a certain period of time.

\setcounter{equation}{0}

\section{A nonlocal Toda type soliton equation}

We begin by collecting the definitions of various important
quantities, cf.~Section~2 in Part~I. We have already recalled
the restrictions on the spectral data $(r,\mu)$ and the
definition of the Cauchy matrix $C(r)$,
cf.~(\ref{ruijsenaars:req1})--(\ref{ruijsenaars:mu}). The dependence on
$\mu$ is
encoded in the
diagonal matrix
\begin{equation}\label{ruijsenaars:defD}
D(r,\mu;x)\equiv \mbox{diag}\, (d(r_1,\mu_1;x),\ldots,
d(r_N,\mu_N;x)),
\end{equation}
where the function $d$ is defined by
\begin{equation}\label{ruijsenaars:defd}
d(\rho,\nu;x)\equiv \left\{
\begin{array}{ll}
\nu(x)e^{-2i\rho x},  &  \mbox{Im}\; \rho \in(0,\pi),
\vspace{1mm}\\
\nu(x)e^{-2i(\rho +i\pi)x},  & \mbox{Im}\; \rho\in(-\pi,0).
\end{array} \right.
\end{equation}
To ease the notation, we often write
\begin{equation}
d_n(x)=d(r_n,\mu_n;x),\qquad n=1,\ldots,N.
\end{equation}

All of the remaining quantities can now be defined in terms of
the solution $R(r,\mu;x)$ to the linear system
\begin{equation}\label{ruijsenaars:sysN}
(D(r,\mu;x)+C(r))R(x)=\zeta,\qquad \zeta\equiv (1,\ldots,1)^t.
\end{equation}
Specifically, introducing the auxiliary functions
\begin{equation}\label{ruijsenaars:la}
\lambda(r,\mu;x)\equiv 1+\sum_{n=1}^Ne^{r_n}R_n(r,\mu;x),
\end{equation}
\begin{equation}\label{ruijsenaars:sig}
\Sigma(r,\mu;x)\equiv \sum_{n=1}^NR_n(r,\mu;x),
\end{equation}
the potentials are given by
\begin{equation}\label{ruijsenaars:Va}
V_a(r,\mu;x)\equiv\lambda(r,\mu;x)/\lambda(r,\mu;x+i),
\end{equation}
\begin{equation}\label{ruijsenaars:Vb}
V_b(r,\mu;x)\equiv\Sigma(r,\mu;x-i)-\Sigma(r,\mu;x),
\end{equation}
and the wave function reads
\begin{equation}\label{ruijsenaars:W}
{\mathcal W}(r,\mu;x,p)\equiv e^{ixp}\left( 1-\sum_{n=1}^N
\frac{R_n(r,\mu;x)}{e^p-e^{-r_n}}\right).
\end{equation}

We now introduce time-dependent multipliers
\begin{equation}\label{ruijsenaars:mut}
\mu_n(r_n;x,t)\equiv
\mu_n(x)\exp\left(it\left[e^{r_n}-e^{-r_n}\right]\right),\qquad
n=1,\ldots,N.
\end{equation}
Correspondingly, the above quantities
(\ref{ruijsenaars:defD})--(\ref{ruijsenaars:W})
are henceforth viewed as depending on time as well. But as a
rule we suppress the dependence on $t$, just as the dependence
on $(r,\mu)$. Until further notice, $t$ may be viewed as a
complex parameter. We denote partial differentiation with
respect to $t$ by a dot. (The factor $i$ in (\ref{ruijsenaars:mut}) occurs
with an eye on later reality restrictions.)

We proceed by obtaining the time derivatives of the above
quantities. The pertinent formulas are far from immediate, so
for clarity we assemble them in a series of propositions.

\begin{prop}
One has
\begin{equation}\label{ruijsenaars:Rd}
\dot{R}(x)=i\, \mbox{\rm diag}
\left(e^{-r_1},\ldots,e^{-r_N}\right)(R(x)-R(x-i))-iV_b(x)R(x).
\end{equation}
\end{prop}

\noindent
{\bf Proof.} Combining (\ref{ruijsenaars:defD}), (\ref{ruijsenaars:defd}) with
(\ref{ruijsenaars:mut}), we obtain
\begin{equation}
\dot{D}(x)=iD_{-}D(x),
\end{equation}
where we have introduced
\begin{equation}
D_{-}\equiv \mbox{diag} \left(e^{r_1}-e^{-r_1},\ldots,e^{r_N}-e^{-r_N}\right).
\end{equation}
Now from (\ref{ruijsenaars:sysN}) we deduce
\begin{equation}
\dot{D}(x)R(x)+(D(x)+C)\dot{R}(x)=0,
\end{equation}
so that
\begin{equation}
(D(x)+C)\dot{R}(x)=-iD_{-}D(x)R(x).
\end{equation}
Therefore, (\ref{ruijsenaars:Rd}) will follow once we show
\begin{equation}\label{ruijsenaars:key}
D_{-}D(x)R(x)=(D(x)+C)\,\mbox{diag}
\left(e^{-r_1},\ldots,e^{-r_N}\right)\!(R(x-i)-R(x))+V_b(x)\zeta.\!\!\!
\end{equation}
(We used (\ref{ruijsenaars:sysN}) to simplify the rhs.)

To prove (\ref{ruijsenaars:key}), consider its $n$th component. Canceling
terms $-\exp(-r_n)d_n(x)R_n(x)$ on the lhs and rhs, it reads
\begin{equation}\arraycolsep=0em
\begin{array}{l}
e^{r_n}d_n(x)R_n(x)    =   d_n(x)e^{-r_n}R_n(x-i)
\vspace{2mm}\\
\displaystyle  \phantom{e^{r_n}d_n(x)R_n(x)   ={}}
+\sum_{j=1}^NC_{nj}e^{-r_j}(R_j(x-i)-R_j(x))+V_b(x).
\end{array}
\end{equation}
In view of (\ref{ruijsenaars:defd}) and (\ref{ruijsenaars:Vb}), this can be
rewritten
as
\begin{equation}\arraycolsep=0em
\begin{array}{l}
e^{r_n}(d_n(x)R_n(x)-d_n(x-i)R_n(x-i))
\vspace{2mm}\\
\displaystyle \qquad =
\sum_{j=1}^N\left(C_{nj}\left(e^{r_n}-\left[e^{r_n}-e^{-r_j}\right]\right)+1\right)
(R_j(x-i)-R_j(x)).
\end{array}
\end{equation}
Now from the definition (\ref{ruijsenaars:Cr}) of the Cauchy matrix we see
that this amounts to
\begin{equation}
d_n(x)R_n(x)-d_n(x-i)R_n(x-i)=\sum_{j=1}^NC_{nj}(R_j(x-i)-R_j(x)).
\end{equation}
By virtue of (\ref{ruijsenaars:sysN}), this is clearly true, so
(\ref{ruijsenaars:key}) follows. \hfill \rule{3mm}{3mm}

\begin{prop}
Introducing the A$\Delta$O
\begin{equation}\label{ruijsenaars:defB}
B\equiv -i(T_i+V_b(x)),
\end{equation}
one has
\begin{equation}\label{ruijsenaars:Wd}
\dot{{\mathcal W}}(x,p)=(B{\mathcal W})(x,p)+ie^p{\mathcal W}(x,p).
\end{equation}
\end{prop}

\noindent
{\bf Proof}. From (\ref{ruijsenaars:W}) we obtain
\begin{equation}
\dot{{\mathcal W}}(x,p)= - e^{ixp}\sum_{n=1}^N
\frac{\dot{R}_n(x)}{e^p-e^{-r_n}}.
\end{equation}
Using (\ref{ruijsenaars:Rd}), this can be rewritten as
\begin{equation}\arraycolsep=0em
\begin{array}{l}
\displaystyle i\dot{{\mathcal W}}(x,p)    =
-e^{ixp}\Sigma(x)+e^pe^{ixp}\sum_{n=1}^N\frac{R_n(x)}{e^p-e^{-r_n}}
\vspace{2mm} \\
\displaystyle \phantom{i\dot{{\mathcal W}}(x,p) ={}}
+e^{ixp}\Sigma(x-i)-e^pe^{ixp}\sum_{n=1}^N\frac{R_n(x-i)}
{e^p-e^{-r_n}}
 -V_b(x)(e^{ixp}-{\mathcal W}(x,p)).
\end{array}
\end{equation}

From the definition (\ref{ruijsenaars:Vb}) of $V_b$ we now see that we can
cancel three terms on the rhs. Then we are left with
\begin{equation}\arraycolsep=0em
\begin{array}{l}
i\dot{{\mathcal W}}(x,p)    =   -e^p({\mathcal W}(x,p)-e^{ixp})
\vspace{2mm}\\
\displaystyle \phantom{i\dot{{\mathcal W}}(x,p) ={}}
  +e^p\left(e^{-p}{\mathcal W}(x-i,p)-e^{ixp}\right)   +V_b(x){\mathcal
W}(x,p),
\end{array}
\end{equation}
which amounts to (\ref{ruijsenaars:Wd}).\hfill \rule{3mm}{3mm}

\medskip

Since the wave function satisfies the A$\Delta$E (\ref{ruijsenaars:Wade}),
and $iB$ equals $A-V_a(x)T_{-i}$ (cf.~(\ref{ruijsenaars:A})), an
alternative formula for the time derivative reads
\begin{equation}
\dot{{\mathcal W}}(x,p)=iV_a(x){\mathcal W}(x+i,p)-ie^{-p}{\mathcal W}(x,p).
\end{equation}
We now proceed with the time derivatives of the potentials.

\begin{prop}
One has
\begin{equation}\label{ruijsenaars:Sigd}
\dot{\Sigma}(x)=i(1-V_a(x)),
\end{equation}
\begin{equation}\label{ruijsenaars:Vbd}
\dot{V_b}(x)=i(V_a(x)-V_a(x-i)).
\end{equation}
\end{prop}

\noindent
{\bf Proof.} In view of (\ref{ruijsenaars:Vb}), the time derivative
(\ref{ruijsenaars:Vbd}) is immediate from (\ref{ruijsenaars:Sigd}). To prove
(\ref{ruijsenaars:Sigd}), we first use the relation
\begin{equation}
{\mathcal W}(x,r_n)=e^{ixr_n}d_n(x)R_n(x),
\end{equation}
and the ${\mathcal W}$-A$\Delta$E (\ref{ruijsenaars:Wade}) to deduce
\begin{equation}
e^{-r_n}R_n(x-i)+e^{r_n}V_a(x)R_n(x+i)+\left(V_b(x)-e^{r_n}-
e^{-r_n}\right)R_n(x)=0.
\end{equation}
Therefore, (\ref{ruijsenaars:Rd}) can be rewritten as
\begin{equation}
i\dot{R}_n(x)=e^{r_n}(R_n(x)-V_a(x)R_n(x+i)),\qquad
n=1,\ldots,N.
\end{equation}
Taking now the sum of these $N$ equations and using
(\ref{ruijsenaars:la})--(\ref{ruijsenaars:Va}), we obtain
\begin{equation}
i\dot{\Sigma}(x)    =
\lambda(x)-1-\frac{\lambda(x)}{\lambda(x+i)}[\lambda(x+i)-1]
  =    V_a(x)-1,
\end{equation}
as asserted. \hfill \rule{3mm}{3mm}

\pagebreak

In order to obtain the time derivative of $V_a(x)$, and also
for later purposes, it is expedient to introduce one more
quantity, namely, the $\tau$-function
\begin{equation}\label{ruijsenaars:tau}
\tau(r,\mu;x,t)\equiv \left|{\bf 1}_N+C(r)D(r,\mu;x,t)^{-1}\right|.
\end{equation}
In view of the alternative representation I(C.30) for
$\lambda(x)$, we readily obtain
\begin{equation}\label{ruijsenaars:la2}
\lambda(x)=\tau (x-i)/\tau(x),
\end{equation}
\begin{equation}\label{ruijsenaars:Va2}
V_a(x)=\tau(x+i)\tau(x-i)/\tau(x)^2.
\end{equation}
We are now prepared for our last proposition.

\begin{prop}
One has
\begin{equation}\label{ruijsenaars:taud}
\dot{\tau}(x)/\tau(x)=-i\Sigma(x),
\end{equation}
\begin{equation}\label{ruijsenaars:Vad}
\dot{V}_a(x)=iV_a(x)(V_b(x+i)-V_b(x)).
\end{equation}
\end{prop}

\noindent
{\bf Proof}. Clearly, (\ref{ruijsenaars:Vad}) follows from
(\ref{ruijsenaars:Va2}),
(\ref{ruijsenaars:Vb}) and (\ref{ruijsenaars:taud}). To prove
(\ref{ruijsenaars:taud}), we first
use (\ref{ruijsenaars:tau}) and Leibniz' rule to write
\begin{equation}
\dot{\tau}(x)=-i\sum_{n=1}^N\tau_n(x).
\end{equation}
Here, $\tau_n(x)$ denotes the determinant of the matrix
obtained from ${\bf 1}_N +CD(x)^{-1}$ when the $n$th column is
replaced by
$\left(e^{r_n}-e^{-r_n}\right)d_n(x)^{-1}(C_{1n},\ldots,C_{Nn})^t$. In
the determinant quotient $\tau_n(x)/\tau(x)$ we now multiply
both matrices from the right by $D(x)$. Then we obtain
$|\Omega_n(x)|/|D(x)+C|$, with $\Omega_n(x)$ the matrix
defined in the paragraph of Part~I containing (C.6). From
I(C.7) we then deduce~(\ref{ruijsenaars:taud}).\hfill\rule{3mm}{3mm}

\medskip

We now turn to some immediate consequences of the above
formulas. First, we have from (\ref{ruijsenaars:A}),
(\ref{ruijsenaars:Vbd}) and
(\ref{ruijsenaars:Vad})
\begin{equation}\label{ruijsenaars:Ad}\arraycolsep=0em
\begin{array}{l}
\dot{A}   =
\dot{V}_a(x)T_{-i}+\dot{V}_b(x)=iV_a(x)(V_b(x+i)-V_b(x))T_{-i}
\vspace{2mm}\\
\phantom{\dot{A}={}}  +i(V_a(x)-V_a(x-i))=    i[V_a(x)T_{-i},T_i+V_b(x)].
\end{array}
\end{equation}
Recalling (\ref{ruijsenaars:defB}), this can be rewritten as a Lax type
equation
\begin{equation}\label{ruijsenaars:Lax}
\dot{A}=[B,A].
\end{equation}
Second, from (\ref{ruijsenaars:taud}), (\ref{ruijsenaars:Sigd}) and
(\ref{ruijsenaars:Va2}) we
deduce
\begin{equation}\label{ruijsenaars:tauto}
\ddot{\tau}(x)\tau(x)-\dot{\tau}(x)^2=\tau(x)^2
-\tau(x+i)\tau(x-i).
\end{equation}
Third, introducing
\begin{equation}\label{ruijsenaars:Psi}
\Psi(r,\mu;x,t)\equiv i\ln (\lambda(r,\mu;x,t))=i\ln
(\tau(r,\mu;x-i,t)/\tau(r,\mu;x,t)),
\end{equation}
we readily obtain
\begin{equation}\label{ruijsenaars:nonto2}
\ddot{\Psi}(x)=i\exp
(i[\Psi(x+i)-\Psi(x)])-i\exp(i[\Psi(x)-\Psi(x-i)]).
\end{equation}

The nonlocal evolution equation (\ref{ruijsenaars:nonto2}) is the Toda
type equation announced in the Introduction,
cf.~(\ref{ruijsenaars:nonto}). Since Lemma~2.2 in Part~I yields
\begin{equation}
\lim_{{\rm Re}\; x \to \infty} \lambda(x)=1,\qquad
\lim_{{\rm Re}\; x \to -\infty} \lambda(x)=\exp\left(2\sum_{n=1}^Nr_n\right),
\end{equation}
it follows from (\ref{ruijsenaars:Psi}) that we have
\begin{equation}
\lim_{{\rm Re}\; x \to \infty} \Psi(x)=0,\qquad
 \lim_{{\rm Re}\; x \to -\infty} \Psi(x)=2i\sum_{n=1}^Nr_n, \qquad
(\mbox{mod}\;{2\pi}).
\end{equation}
The multi-valuedness indicated here is inevitably present when
we let $x$ and $t$ vary over~${\mathbb C}$. Indeed, $\tau(x,t)$
(\ref{ruijsenaars:tau})  has zeros in general (as well as poles whenever
$\mu(x)$ is non-constant). Therefore,
$\Psi(x,t)$ has logarithmic branch points. Note, however, that
this multi-valuedness is of no consequence in (\ref{ruijsenaars:nonto2}).

We continue by studying reality restrictions. Specifically, we
ask first: Can one choose the spectral data $(r,\mu)$ such
that $\Psi(r,\mu;x,t)$ is real-valued for real $x$ and $t$?

This question can be answered in the affirmative by using the
results of Appendix~D in Part~I. We showed there that the
A$\Delta$O $A$ (\ref{ruijsenaars:A}) is formally self-adjoint on
$L^2({\mathbb R},dx)$
whenever $r_1,\ldots,r_N$ are purely imaginary and the
functions $i\exp(-r_n)\mu_n(x)$, $n=1,\ldots,N$, are real-valued
for real $x$. Along the way, we obtained as another
consequence of these restrictions the relation
\begin{equation}
\overline{\lambda(\overline{x})}=1/\lambda(x),\qquad x\in {\mathbb C},
\end{equation}
cf.~I(D.17).
Imposing these restrictions and choosing $t$ real from now on,
the time-dependent factors $\exp (it[\exp(r_n)-\exp(-r_n)])$
belong to
$(0,\infty)$, so we deduce
\begin{equation}\label{ruijsenaars:phase}
\overline{\lambda(x,t)}=1/\lambda(x,t),\qquad x,t\in {\mathbb R}.
\end{equation}

Now $\lambda(x,t)$ is meromorphic in $x$ and $t$. For $x$, $t$
real, (\ref{ruijsenaars:phase}) entails that
$\lambda(x,t)$ is a phase factor, so in particular no zeros or
poles occur. As a consequence, we need only fix the logarithm
branch in (\ref{ruijsenaars:Psi}) by requiring
\begin{equation}\label{ruijsenaars:branch}
\lim_{x\to\infty} \Psi(x,t)=0,\qquad t\in {\mathbb R},
\end{equation}
to obtain a real-valued, real-analytic function
\begin{equation}
\Psi\, :\, {\mathbb R}^2\to {\mathbb R},\qquad (x,t)\mapsto \Psi(x,t).
\end{equation}
Keeping $t$ real, its (multi-valued) continuation to complex
$x$ satisfies
\begin{equation}
\overline{\Psi(t,\overline{x})}=\Psi(t,x),\qquad (\mbox{mod}\; {2\pi}).
\end{equation}

Let us now consider the characteristics of the solutions to
(\ref{ruijsenaars:nonto2}) with the above reality restrictions in force,
i.e.,
\begin{equation}
\mbox{Re}\; r_n=0,\qquad n=1,\ldots,N,
\end{equation}
\begin{equation}
\mbox{Re} \left(e^{-r_n}\mu_n(x)\right)=0,\qquad
 x\in{\mathbb R},\qquad n=1,\ldots,N.
\end{equation}
Taking first $N=1$, and choosing
$r_1=i\kappa^{+}$, $\kappa^{+}\in(0,\pi)$, we get from (\ref{ruijsenaars:tau})
and (\ref{ruijsenaars:Psi})
\begin{equation}
\tau(x,t)=1+(2i\sin \kappa^{+})^{-1}\exp (-2\kappa^{+} x+2t\sin \kappa^{+}
)\mu_1(x)^{-1},
\end{equation}
\begin{equation}
\Psi(x,t)=i\ln \left(
\frac{2ie^{-i\kappa^{+}}\mu_1(x)\sin \kappa^{+} +
e^{i\kappa^{+}}\exp(-2\kappa^{+} x+2t\sin\kappa^{+})}
{2ie^{-i\kappa^{+}}\mu_1(x)\sin \kappa^{+} +
e^{-i\kappa^{+}}\exp(-2\kappa^{+} x+2t\sin\kappa^{+})}
\right) .
\end{equation}
Since the function $ie^{-i\kappa^{+}}\mu_1(x)$ is real-valued for
real $x$, this indeed yields a real-valued function for
$(x,t)\in{\mathbb R}^2$. But whenever $\mu_1(x)$ is non-constant,
$\Psi(x,t)$ cannot be viewed as a 1-soliton solution. Indeed, in
that case it is not of the traveling wave form $f(x-vt)$.

Choosing however
\begin{equation}\label{ruijsenaars:mu+}
2ie^{-i\kappa^{+}}\mu_1(x)\sin\kappa^{+} =\exp(-2\kappa^{+}
a^{+}),\qquad a^{+}\in
{\mathbb R},
\end{equation}
we do get a function that is not only of the form $f(x-vt)$,
but also of the kink type, in the sense that its
$x$-derivative is positive and exponentially localized around
its maximum at $x=a^{+}+vt$:
\begin{equation}\label{ruijsenaars:der+}
\Psi^{\prime}(x,t)=\frac{2\kappa^{+}\sin\kappa^{+}}{\cos\kappa^{+}+\cosh
2\kappa^{+}(x-a^{+}-v(\kappa^{+})t)},\qquad
v(\kappa )\equiv \frac{\sin \kappa }{\kappa }.
\end{equation}
It is readily checked that the velocity function $v(\kappa )$
decreases monotonically from~1 to~0 as $\kappa$ goes from~0
to~$\pi$. Observe also that one has
\begin{equation}\label{ruijsenaars:psi+as}
\lim_{x\to -\infty}\Psi(x,t)=-2\kappa^{+},\qquad t\in{\mathbb R}.
\end{equation}

Consider next the choice $r_1=i\kappa^{-}-i\pi$, $\kappa^{-}\in(0,\pi)$.
Then we obtain
\begin{equation}
\tau(x,t)=1-(2i\sin \kappa^{-})^{-1}\exp (-2\kappa^{-} x-2t\sin
\kappa^{-} )\mu_1(x)^{-1},
\end{equation}
\begin{equation}
\Psi(x,t)=i\ln \left(
\frac{-2ie^{-i\kappa^{-}}\mu_1(x)\sin \kappa^{-} +
e^{i\kappa^{-}}\exp(-2\kappa^{-} x-2t\sin\kappa^{-})}
{-2ie^{-i\kappa^{-}}\mu_1(x)\sin \kappa^{-} +
e^{-i\kappa^{-}}\exp(-2\kappa^{-} x-2t\sin\kappa^{-})}
\right) .
\end{equation}

Once again, we need $\mu_1(x)$ to be constant for this
function to have 1-soliton characteristics. Setting
\begin{equation}\label{ruijsenaars:mu-}
-2ie^{-i\kappa^{-}}\mu_1(x)\sin\kappa^{-} =\exp(-2\kappa^{-}
a^{-}),\qquad a^{-}\in
{\mathbb R},
\end{equation}
we now obtain
\begin{equation}\label{ruijsenaars:der-}
\Psi^{\prime}(x,t)=\frac{2\kappa^{-}\sin\kappa^{-}}
{\cos\kappa^{-}+\cosh
2\kappa^{-}(x-a^{-}+v(\kappa^{-})t)}.
\end{equation}
Thus, $\Psi(x,t)$ is a 1-kink solution moving to the left with
speed in the interval $(0,1)$. Clearly, it satisfies
\begin{equation}\label{ruijsenaars:psi-as}
\lim_{x\to -\infty}\Psi(x,t)=-2\kappa^{-},\qquad t\in{\mathbb R}.
\end{equation}

Proceeding with the arbitrary-$N$ case, we can clearly make
choices for $\mu_n(x)$ corresponding to (\ref{ruijsenaars:mu+}) and
(\ref{ruijsenaars:mu-}), depending on whether $r_n$ belongs to $i(0,\pi)$
or $i(-\pi,0)$. Doing so, we obtain real-valued,
real-analytic solutions. In Section~6 we study these solutions
in detail, demonstrating in particular that they may be viewed
as $N$-soliton solutions.

\setcounter{equation}{0}
\section{Reflectionless self-adjoint Schr\"odinger operators}

In this section we clarify the relation of the above A$\Delta$Os
and their reflectionless eigenfunctions to the reflectionless
self-adjoint Schr\"odinger operators considered in the IST
framework~\cite{ruijsenaars:scot,ruijsenaars:cade,ruijsenaars:newe}. Thus,
we start from
a continuous real-valued potential $V_H(x)$ with decay at $\pm
\infty$ given by
\begin{equation}
V_H(x)=O\left(|x|^{-e}\right),\qquad x\to \pm \infty ,
\end{equation}
with a suitable positive exponent $e$. Then it is clear that
the Schr\"odinger operator on $L^2({\mathbb R},dx)$ given by
\begin{equation}
(Hf)(x)\equiv -f'' (x)+V_H(x)f(x),
\end{equation}
 is self-adjoint on the natural domain of the
free Hamiltonian $-d^2/dx^2$.

The reflectionless operators can now be characterized by the
existence of an $H$-eigen\-func\-tion
\begin{equation}
(H{\mathcal W}_H)(x,p)=p^2{\mathcal W}_H(x,p),
\end{equation}
with asymptotics
\begin{equation}
{\mathcal W}_H(x,p)\sim \left\{
\begin{array}{ll}
\exp (ixp),  &  x\to\infty, \vspace{1mm}\\
a_H(p)\exp (ixp),  &  x\to -\infty .
\end{array}
\right.
\end{equation}
The IST framework yields a complete classification of such
operators: The function $a_H(p)$ is of the form
\begin{equation}\label{ruijsenaars:aH}
a_H(p)=\prod_{n=1}^N\frac{p-i\kappa_n}{p+i\kappa_n},\qquad
0<\kappa_N<\cdots <\kappa_1,
\end{equation}
and for each such $S$-matrix there exists an
$N$-dimensional family of potentials parametrized by
normalization coefficients $\nu_1,\ldots,\nu_N\in
(0,\infty)$. The eigenfunction ${\mathcal W}_H(x,p)$ yields bound states
\begin{equation}\label{ruijsenaars:bsH}
\phi_n(\cdot)\equiv {\mathcal W}_H(\cdot,i\kappa_n),\qquad
n=1,\ldots,N,
\end{equation}
satisfying
\begin{equation}
\int_{-\infty}^\infty |\phi_n(x)|^2dx =1/\nu_n,\qquad n=1,\ldots,N.
\end{equation}
Thus $H$ has continuous spectrum $[0,\infty)$ and
discrete spectrum $-\kappa_1^2,\ldots,-\kappa_N^2$. The bound
states are pairwise orthogonal, and the improper eigenfunction
${\mathcal W}_H(x,p),p\in {\mathbb R}$, gives rise to the (Schwartz) kernel of
an isometry from $L^2({\mathbb R},dp)$ onto the orthocomplement of the
bound state subspace.

As we will show in Part~III, the state of affairs concerning
Hilbert space properties of the above A$\Delta$Os and their
reflectionless eigenfunctions deviates from this ``Schr\"odinger
scenario'' in several ways. In this section, however, we only
aim to show how the self-adjoint reflectionless operators $H$
and their eigenfunctions ${\mathcal W}_H(x,p)$ arise as limits of our
A$\Delta$Os and their reflectionless eigenfunctions.

Turning to the details, we fix
\begin{equation}
(\kappa,\nu)\in(0,\infty)^{2N},\qquad
0<\kappa_N<\cdots<\kappa_1.
\end{equation}
Then we choose
\begin{equation}
r_n=i\beta\kappa_n,\qquad \beta\in (0,\pi/\kappa_1),\qquad
n=1,\ldots,N,
\end{equation}
and constant multipliers
\begin{equation}
\mu_n(x)=-ie^{r_n}/\beta\nu_n,\qquad n=1,\ldots,N.
\end{equation}
This entails that the A$\Delta$O $A$ (\ref{ruijsenaars:A}) is formally
self-adjoint, cf.~Theorem~D.1 in Part~I. Moreover, since all
numbers $r_1,\ldots,r_N$ have imaginary part in $(0,\pi)$, we
may invoke Theo\-rem~3.3 in Part~I. It entails in particular that
$A$ can be rewritten as
\begin{equation}
A=S_{+}^2-2,
\end{equation}
where
\begin{equation}
S_{+}=T_{i/2}+V(x)T_{-i/2},
\end{equation}
\begin{equation}
V(x)=\sum_{n=1}^N(R_n(x-i/2)-R_n(x))+1.
\end{equation}
Furthermore, the wave function is an $S_{+}$-eigenfunction:
\begin{equation}
(S_{+}{\mathcal W})(x,p)=\left(e^{p/2}+e^{-p/2}\right){\mathcal W}(x,p).
\end{equation}

We now introduce a scaling $x\to \beta^{-1}x$, $p\to\beta p$ in the
various quantities at hand, and study the $\beta\to 0$ limit.
Specifically, introducing first
\begin{equation}
R_{\beta,n}(x)\equiv \beta^{-1}R_n\left(\beta^{-1}x\right),
\end{equation}
the system (\ref{ruijsenaars:sysN}) with $x\to \beta^{-1}x$ can be rewritten
as
\begin{equation}\arraycolsep=0em
\begin{array}{l}
\displaystyle \frac{\exp(i\beta\kappa_n+2\kappa_nx)}{i\nu_n}R_{\beta,n}(x)
\vspace{3mm}\\
\displaystyle \qquad
+\sum_{j=1}^N\frac{\beta}{\exp(i\beta\kappa_n)-\exp(-i\beta
\kappa_j)}R_{\beta,j}(x)=1,\qquad  n=1,\ldots,N.
\end{array}
\end{equation}

From this one readily deduces that $R_{\beta,n}(x)$ is
holomorphic at $\beta=0$, with a limit
\begin{equation}\label{ruijsenaars:RH}
\lim_{\beta\to 0}R_{\beta,n}(x)\equiv R_n^H(x),\qquad
n=1,\ldots,N,
\end{equation}
that solves the system
\begin{equation}\label{ruijsenaars:sysH}
\frac{\exp(2\kappa_nx)}{i\nu_n}R_n^H(x)+
\sum_{j=1}^N\frac{1}{i\kappa_n+i\kappa_j}R_j^H(x)=1,\qquad
n=1,\ldots,N.
\end{equation}

Secondly, setting
\begin{equation}
{\mathcal W}_{\beta}(x,p)\equiv {\mathcal W}\left(\beta^{-1}x,\beta p\right),
\end{equation}
it follows that ${\mathcal W}_{\beta}(x,p)$ is holomorphic at $\beta=0$
as well, with a limit
\begin{equation}\label{ruijsenaars:Wlim}
\lim_{\beta\to 0}{\mathcal W}_{\beta}(x,p)=e^{ixp}\left(
1-\sum_{j=1}^N\frac{R_j^H(x)}{p+i\kappa_j}\right) \equiv
{\mathcal W}_H(x,p).
\end{equation}

Thirdly, introducing
\begin{equation}
S_{\beta,+}\equiv \exp
(-i\beta\partial_x/2)+V_{\beta}(x)\exp(i\beta\partial_x/2),
\end{equation}
with
\begin{equation}
V_{\beta}(x)\equiv \beta \sum_{n=1}^N
(R_{\beta,n}(x-i\beta/2)-R_{\beta,n}(x))+1,
\end{equation}
we clearly have
\begin{equation}
S_{\beta,+}{\mathcal W}_{\beta}(x,p)=\left(e^{\beta p/2}+e^{-\beta
p/2}\right){\mathcal W}_{\beta}(x,p).
\end{equation}
Now from (\ref{ruijsenaars:RH}) we deduce
\begin{equation}
V_{\beta}(x)=1+(\beta/2)^2V_H(x)+O\left(\beta^3\right),\qquad \beta\to 0,
\end{equation}
where
\begin{equation}\label{ruijsenaars:VH}
V_H(x)\equiv -2i\sum_{n=1}^N\partial_xR_n^H(x).
\end{equation}
As a consequence, we obtain
\begin{equation}
S_{\beta,+}=2+(\beta/2)^2H+O\left(\beta^3\right),\qquad \beta\to 0,
\end{equation}
with
\begin{equation}
H=-\partial_x^2+V_H(x),
\end{equation}
and
\begin{equation}
(H{\mathcal W}_H)(x,p)=p^2 {\mathcal W}_H(x,p).
\end{equation}

Proceeding in this way, we have actually obtained all of the
reflectionless self-adjoint Schr\"odinger operators delineated
at the beginning of this section. To transform them to a more
familiar form, we need only invoke (\ref{ruijsenaars:Wlim}) and the system
(\ref{ruijsenaars:sysH}) to write
\begin{equation}\arraycolsep=0em
\begin{array}{l}
\displaystyle \phi_n(x)\equiv {\mathcal W}_H(x,i\kappa_n)    =
e^{-\kappa_nx}\left(
1-\sum_{j=1}^N\frac{R_n^H(x)}{i\kappa_n+i\kappa_j}\right)
\vspace{2mm}\\
\displaystyle \phantom{\phi_n(x)\equiv{}}   =
\frac{e^{\kappa_nx}}{i\nu_n}R_n^H(x),\qquad
n=1,\ldots,N.
\end{array}
\end{equation}
Thus $V_H(x)$ (\ref{ruijsenaars:VH}) can be rewritten as
\begin{equation}
V_H(x)=2\sum_{n=1}^N\nu_n\partial_x\left(e^{-\kappa_nx}\phi_n(x)\right).
\end{equation}
This is the well-known formula expressing the reflectionless
potentials in terms of their bound states
$\phi_1(x),\ldots,\phi_N(x)$.

To conclude this section, we add three remarks. First, we point
out that the asymptotics
\begin{equation}
{\mathcal W}(x,p)\sim \left\{
\begin{array}{ll}
\exp (ixp),  &  x\to\infty, \vspace{1mm}\\
a(p)\exp (ixp),  &  x\to -\infty ,
\end{array}
\right.
\end{equation}
\begin{equation}
a(p)=\prod_{n=1}^N\frac{e^p-e^{r_n}}{e^p-e^{-r_n}},
\end{equation}
derived in Theorem~2.3 of Part~I yields the function $a_H(p)$
(\ref{ruijsenaars:aH}) when one substitutes $p\to \beta p$,
$r_n\to i\beta\kappa_n$, and takes $\beta\to 0$.
Second, we observe that we could also have used the
A$\Delta$O
\begin{equation}
A_{\beta}\equiv S_{\beta,+}^2-2
\end{equation}
to arrive at the Schr\"odinger operator $H$, as will be clear
from the above.

Finally, just as the operator $H$ may be viewed as a
reduced Hamiltonian for a Galilei-invariant two-particle
system, one may view the A$\Delta$O $S_{\beta,+}$ (or $A_{\beta}$)
as the reduced Hamiltonian for a Poincar\'e-invariant
two-particle system. In this scenario the limit $\beta\to 0$
amounts to the nonrelativistic limit, cf.~Ref.~\cite{ruijsenaars:posen}.

\setcounter{equation}{0}
\section{Reflectionless self-adjoint Jacobi operators\\
  and Toda lattice solitons}

In this section we show how the reflectionless self-adjoint
Jacobi operators and Toda lattice solitons arise via analytic
continuation and $x$-discretization. For comparison purposes,
especially useful are the monographs
Refs.~\cite{ruijsenaars:toda,ruijsenaars:fata}
and Flaschka's paper Ref.~\cite{ruijsenaars:flas}. For the Jacobi operator
\begin{equation}\label{ruijsenaars:Jac}
(Jf)(n)=a(n-1)f(n-1)+a(n)f(n+1)+b(n)f(n)
\end{equation}
on $l^2({\mathbb Z})$ one requires decay
\begin{equation}\label{ruijsenaars:abas}
a(n)=1/2 +O\left(|n|^{-e_a}\right),\qquad b(n)=O\left(|n|^{-e_b}\right),\qquad
n\to\pm\infty,
\end{equation}
for suitable positive exponents $e_a$, $e_b$. Requiring in
addition that $a(n)$, $b(n)$ be real, it is clear that $J$ is a
bounded self-adjoint operator on $l^2({\mathbb Z})$.

With these requirements in effect, the reflectionless Jacobi
operators are characterized by the existence of a
$J$-eigenfunction ${\mathcal W}_J(n,p)$ satisfying
\begin{equation}\label{ruijsenaars:WJ}
(J{\mathcal W}_J)(n,p)=\cos(p){\mathcal W}_J(n,p),
\end{equation}
with asymptotics
\begin{equation}\label{ruijsenaars:WJ+}
{\mathcal W}_J(n,p)\sim \exp(inp),\qquad n\to\infty,
\end{equation}
\begin{equation}\label{ruijsenaars:WJ-}
{\mathcal W}_J(n,p)\sim a_J(p)\exp(inp),\qquad n\to -\infty.
\end{equation}

From the IST formalism it follows that the function $a_J(p)$ is
of the form
\begin{equation}\label{ruijsenaars:aJ}
a_J(p)=\prod_{j=1}^{N_{+}}\frac{\sin [(p-i\kappa^{+}_j)/2]}
{\sin [(p+i\kappa^{+}_j)/2]}\cdot
\prod_{l=1}^{N_{-}}\frac{\cos [(p-i\kappa^{-}_l)/2]}
{\cos [(p+i\kappa^{-}_l)/2]},
\end{equation}
where $0<\kappa^{\delta}_{N_{\delta}}<\cdots <\kappa_1^{\delta}$,
$\delta =+,-$. For each $a_J(p)$ there exists an
$(N_{+}+N_{-})$-dimensional family of Jacobi operators,
parametrized by positive normalization coefficients
$\nu_1^{\delta},\ldots,\nu_{N_{\delta}}^{\delta},\delta =+,-$.
The operator $J$ has $N_{+}+N_{-}$ bound states, given by
\begin{equation}\arraycolsep=0em
\begin{array}{l}
\phi_j^{+}(n)={\mathcal W}_J(n,i\kappa_j^{+}),\qquad
 j=1,\ldots,N_{+},
\vspace{1mm}\\
\ \phi_l^{-}(n)={\mathcal W}_J(n,\pi +i\kappa_l^{-}),\qquad
l=1,\ldots,N_{-},
\end{array}
\end{equation}
and normalized as
\begin{equation}
\sum_{n\in{\mathbb Z}}\left|\phi_k^{\delta}(n)\right|^2=1/\nu_k^{\delta},\qquad
k=1,\ldots,N_{\delta},\qquad \delta=+,-.
\end{equation}
The bound state energies equal $\delta \cosh
(\kappa_j^{\delta})$, $j=1,\ldots,N_{\delta}$, $\delta=+,-$.

Fixing the above spectral data, we choose
\begin{equation}
N\equiv N_{+}+N_{-},
\end{equation}
and let
\begin{equation}
r_j=e^{i\eta}\kappa^{+}_j,\qquad j=1,\ldots,N_{+},\qquad
r_{N_{+}+l}=e^{i\eta}\kappa_l^{-}-i\pi,\qquad l=1,\ldots,N_{-}.
\end{equation}
Here, we choose $\eta\in(0,\eta_s)$, with $\eta_s\in(0,\pi/2)$
satisfying $\kappa_1^{\delta}\sin \eta_s<\pi$, $\delta=+,-$. Then the
requirements (\ref{ruijsenaars:req1}) and (\ref{ruijsenaars:req2}) are
clearly met. We
also choose constant multipliers
\begin{equation}\arraycolsep=0em
\begin{array}{l}
\mu_j(x)=\exp(r_j)/\nu_j^{+},\qquad j=1,\ldots,N_{+},
\vspace{1mm}\\
\mu_{N_{+}+l}(x)=\exp(r_{N_{+}+l})/\nu_l^{-},\qquad
l=1,\ldots,N_{-}.
\end{array}
\end{equation}

Next, we substitute
\begin{equation}
x\to i e^{-i\eta} n ,\qquad p\to -ie^{i\eta}p,\qquad n\in{\mathbb Z},\qquad
\eta\in(0,\eta_s),
\end{equation}
in the above quantities, and study the $\eta\to 0$ limit. To
this end, it is convenient (both for notation and for
comparison purposes) to introduce further parameters
\begin{equation}\label{ruijsenaars:znu}
z_j\equiv \left\{
\begin{array}{l}
\exp (-\kappa_j^{+}), \vspace{1mm}\\
-\exp(-\kappa^{-}_{j-N_{+}}),
\end{array}
\right.
\qquad \nu_j\equiv
\left\{
\begin{array}{ll}
\nu_j^{+}, &   j=1,\ldots,N_{+},
\vspace{1mm}\\
\nu^{-}_{j-N_{+}},  &  j=N_{+}+1,\ldots,N.
\end{array}
\right.
\end{equation}
Employing these parameters, one easily checks
\begin{equation}\label{ruijsenaars:Dlim}
\lim_{\eta\to 0} D\left(ie^{-i\eta} n\right)=\mbox{diag}
\left(1/z_1^{2n+1}\nu_1,\ldots,1/z_N^{2n+1}\nu_N\right)\equiv D^J(n),
\end{equation}
\begin{equation}
\lim_{\eta\to 0} C_{jk}=\frac{z_j}{1-z_jz_k}\equiv C_{jk}^J/z_k,\qquad
j,k=1,\ldots,N.
\end{equation}
Using $z_j\in(-1,1)$, the matrix $C^J$ is readily seen to be
positive. (This can be deduced from Cauchy's identity.
Alternatively, following Flaschka~\cite{ruijsenaars:flas}, one need only
write $1/(1-z_jz_k)$ as a geometric series to verify
positivity.)

As a consequence, the matrix
\begin{equation}\label{ruijsenaars:LJ}
L^J(n)\equiv\lim_{\eta\to 0} CD\left(ie^{-i\eta}n\right)^{-1},\qquad
n\in{\mathbb Z},
\end{equation}
is given by
\begin{equation}\label{ruijsenaars:LCJ}
L^J(n)_{jk}=C_{jk}^Jz_k^{2n}\nu_k,\qquad j,k=1,\ldots,N,\qquad
n\in{\mathbb Z}.
\end{equation}
Therefore, $L^J(n)$ is similar to a positive matrix, so ${\bf
1}_N+L^J(n)$ is invertible. From this and the limit
(\ref{ruijsenaars:Dlim}) it readily follows that
\begin{equation}
\lim_{\eta\to 0} R_j\left(ie^{-i\eta} n\right)\equiv R_j^J(n)
\end{equation}
exists for all $n\in{\mathbb Z}$. Since $R_j^J(n)$ solves a system
\begin{equation}\label{ruijsenaars:sysJ}
\left(z_j^{2n+1}\nu_j\right)^{-1}R_j^J(n)+\sum_{k=1}^N\frac{z_j}
{1-z_jz_k}R_k^J(n)=1,\qquad j=1,\ldots,N,
\end{equation}
with real coefficients, it is real, too.

Next, one easily verifies
\begin{equation}
\lim_{\eta\to 0} D\left(ie^{-i\eta} n \pm i\right)=D^J(n\pm i),\qquad n\in
{\mathbb Z}.
\end{equation}

From this it is not hard to deduce
\begin{equation}
\lim_{\eta\to 0} R_j\left(ie^{-i\eta}n \pm i\right)= R_j^J(n\pm i),\qquad
j=1,\ldots,N,\qquad
n\in {\mathbb Z}.
\end{equation}
(Indeed, the limit of the system (\ref{ruijsenaars:sysN}) in small complex
neighborhoods of $x=in\pm i$ can be controlled by exploiting
(\ref{ruijsenaars:sysJ}) with $n\to n\pm 1$.)

Using the above, the remaining pertinent limits can be easily
obtained. First, recalling (\ref{ruijsenaars:tau})--(\ref{ruijsenaars:Va2})
(with
$t=0$), one gets
\begin{equation}\label{ruijsenaars:tauJ}
\lim_{\eta\to 0} \tau \left(ie^{-i\eta}n\right)=\left|{\bf
1}_N+L^J(n)\right|\equiv \tau^J(n),
\end{equation}
\begin{equation}\label{ruijsenaars:laJ}
\lim_{\eta\to 0}
\lambda\left(ie^{-i\eta}n\right)=\tau^J(n-1)/\tau^J(n)\equiv\lambda^J(n),
\end{equation}
\begin{equation}\label{ruijsenaars:an}
\lim_{\eta\to 0}
V_a\left(ie^{-i\eta}n\right)=\lambda^J(n)/\lambda^J(n+1)\equiv(2a(n))^2.
\end{equation}
This defines $a(n)\in(0,\infty)$, since
(\ref{ruijsenaars:tauJ}) entails $\tau^J(n)>0$.
Secondly, (\ref{ruijsenaars:Vb}) yields
\begin{equation}\label{ruijsenaars:bn}
\lim_{\eta\to 0}
V_b\left(ie^{-i\eta}n\right)=\sum_{j=1}^N\left(R_j^J(n-1)-R_j^J(n)\right)\equiv
2b(n).
\end{equation}
Since $R_j^J(n)$ is real, $b(n)$ is real, too.

Thirdly, consider the wave function
\begin{equation}
{\mathcal W}_{\eta}(x,p)\equiv {\mathcal
W}\left(ie^{-i\eta}x,-ie^{i\eta}p\right).
\end{equation}
Its limit
\begin{equation}\label{ruijsenaars:WhJ}
\lim_{\eta\to 0} {\mathcal W}_{\eta}(n,p)=e^{inp}\left(
1-\sum_{j=1}^N
\frac{R_j^J(n)}{e^{-ip}-z_j}\right) \equiv \hat{{\mathcal W}}^J(n,p),\qquad
n\in{\mathbb Z},
\end{equation}
is once again immediate from the above.
Now when we fix $n\in{\mathbb Z}$, the A$\Delta$E (\ref{ruijsenaars:Wade})
yields
\begin{equation}\arraycolsep=0em
\begin{array}{l}
{\mathcal W}_{\eta}
\left(n-e^{i\eta},p\right)+V_a\left(ie^{-i\eta}n\right){\mathcal
W}_{\eta}\left(n+e^{i\eta},p\right)
\vspace{1mm}\\
\qquad +\left(V_b\left(ie^{-i\eta}n\right)-2\cos
\left(e^{i\eta}p\right)\right){\mathcal W}_{\eta}(n,p)=0.
\end{array}
\end{equation}
Taking $\eta$ to 0, we deduce that $\hat{{\mathcal W}}^J(n,p)$ satisfies the
discrete difference equation
\begin{equation}\label{ruijsenaars:Whde}
\hat{{\mathcal W}}^J(n-1,p)+4a(n)^2\hat{{\mathcal W}}^J(n+1,p)+2b(n)
\hat{{\mathcal W}}^J(n,p)=2\cos (p)\hat{{\mathcal W}}^J(n,p).
\end{equation}
This is not yet of the Jacobi form (\ref{ruijsenaars:WJ}),
cf.~(\ref{ruijsenaars:Jac}),
and we now proceed to explain how the connection is to be made.

Our reasoning involves the $|n|\to\infty$ asymptotics of the
above quantities. This asymptotics does not follow from our
previous work. For one thing, we have taken a limit that goes
beyond the parameter regime studied in Part~I. But even when
one ignores this, the pertinent asymptotics is quite different
from the one in Part~I. Indeed, here we need the asymptotics in
the direction of the shifts $n\to n\pm 1$, whereas in the
A$\Delta$O context we are dealing with the asymptotics $\mbox{Re} \;x\to
\pm\infty$, which is orthogonal to the shifts $x\to x\pm i$
in~(\ref{ruijsenaars:A}).

It is however a simple matter to obtain the desired
$n\to\infty$ asymptotics directly. Indeed, from (\ref{ruijsenaars:sysJ})
and (\ref{ruijsenaars:znu}) one readily deduces
\begin{equation}\label{ruijsenaars:RJ+}
R_j^J(n)=O(\exp(-\rho n)),\qquad n\to\infty,\quad
j=1,\ldots,N,\quad  \rho \equiv
2\min (\kappa_{N_{+}}^{+},\kappa_{N_{-}}^{-}).
\end{equation}
Likewise, from (\ref{ruijsenaars:LCJ}) one has
\begin{equation}
L^J(n)_{jk}=O(\exp(-\rho n)),\qquad n\to\infty,\qquad
j,k=1,\ldots,N.
\end{equation}
Then (\ref{ruijsenaars:tauJ}) yields
\begin{equation}
\tau^J(n)=1+O(\exp(-\rho n)),\qquad n\to\infty,
\end{equation}
so from (\ref{ruijsenaars:laJ}) and (\ref{ruijsenaars:an}) one obtains
\begin{equation}\label{ruijsenaars:laJ+}
\lambda^J(n)=1+O(\exp(-\rho n)),\qquad n\to\infty,
\end{equation}
\begin{equation}\label{ruijsenaars:an+}
2a(n)=1+O(\exp(-\rho n)),\qquad n\to\infty.
\end{equation}
Hence the infinite product
\begin{equation}\label{ruijsenaars:rhon}
\Pi(n)\equiv \prod_{m=n}^{\infty}(2a(m))^{-1}
\end{equation}
converges, and one has
\begin{equation}
\Pi(n)=\lambda^J(n)^{-1/2}.
\end{equation}

When we now renormalize $\hat{{\mathcal W}}^J(n,p)$ by introducing
\begin{equation}\label{ruijsenaars:WJ2}
{\mathcal W}^J(n,p)\equiv \lambda^J(n)^{-1/2}\hat{{\mathcal W}}^J(n,p),\qquad
n\in{\mathbb Z},
\end{equation}
then it follows from (\ref{ruijsenaars:Whde}) that we have
\begin{equation}\label{ruijsenaars:Wde}
a(n-1){\mathcal W}^J(n-1,p)+a(n){\mathcal W}^J(n+1,p)+b(n){\mathcal W}^J(n,p)
=\cos(p){\mathcal W}^J(n,p),
\end{equation}
which is of the Jacobi form (\ref{ruijsenaars:WJ}).
It is therefore clear from the above that the $n\to\infty$
asymptotics is in accord with (\ref{ruijsenaars:abas}) and
(\ref{ruijsenaars:WJ+}): From
(\ref{ruijsenaars:RJ+}) and (\ref{ruijsenaars:bn}) one gets
\begin{equation}
b(n)=O(\exp (-\rho n)),\qquad n\to\infty,
\end{equation}
which, together with (\ref{ruijsenaars:an+}), agrees with
(\ref{ruijsenaars:abas}) for
$n\to\infty$. Moreover, (\ref{ruijsenaars:RJ+}), (\ref{ruijsenaars:WhJ}) and
(\ref{ruijsenaars:an+})--(\ref{ruijsenaars:WJ2}) entail
(\ref{ruijsenaars:WJ+}).

It remains to show that the asymptotics for $n\to -\infty$
works out as announced. This involves a little more work. Let
us first note that (\ref{ruijsenaars:sysJ}) entails that $R^J(n)$ has a
finite limit $R^J(-\infty)$ for $n\to -\infty$, satisfying
\begin{equation}
\sum_{k=1}^N\frac{z_j}{1-z_jz_k}R_k^J(-\infty)=1,\qquad
j=1,\ldots,N.
\end{equation}
Using Cramer's rule, this can be improved to
\begin{equation}
R_j^J(n)=R_j^J(-\infty)+O(\exp(\rho n)),\qquad n\to -\infty,\qquad
j=1,\ldots,N.
\end{equation}
Therefore, (\ref{ruijsenaars:bn}) yields
\begin{equation}
b(n)=O(\exp(\rho n)),\qquad n\to -\infty.
\end{equation}

Next, from (\ref{ruijsenaars:laJ}), (\ref{ruijsenaars:tauJ}) and
(\ref{ruijsenaars:LCJ}) we have
\begin{equation}\label{ruijsenaars:la-}\arraycolsep=0em
\begin{array}{l}
\displaystyle \lambda^J(n)    =
\frac{\left|\mbox{diag} \left(z_1^{-2n},\ldots
,z_N^{-2n}\right)+\left(C_{jk}^Jz_k^{-2}\nu_k\right)\right|}
{\left|\mbox{diag}\left(z_1^{-2n},\ldots
,z_N^{-2n}\right)+\left(C_{jk}^J\nu_k\right)\right|}
\vspace{3mm}\\
\displaystyle \phantom{\lambda^J(n)}
    =    \prod_{k=1}^Nz_k^{-2}+O(\exp(\rho n)), \qquad n\to
-\infty.
\end{array}
\end{equation}
Therefore, we obtain from (\ref{ruijsenaars:an})
\begin{equation}
a(n)=1/2+O(\exp (\rho n)),\qquad n\to -\infty.
\end{equation}
Summarizing, the asymptotics for $|n|\to\infty$ of the
coefficients $a(n)$ and $b(n)$ agrees with~(\ref{ruijsenaars:abas}).

We are left with determining the $n\to -\infty$ asymptotics of
the wave function. A quick way to obtain this is to invoke the
alternative representation for ${\mathcal W}(x,p)$ from Theorem~C.3 of
Part~I. It entails that $\hat{{\mathcal W}}^J(n,p)$ (\ref{ruijsenaars:WhJ})
can also
be written
\begin{equation}\label{ruijsenaars:Wh2}
\hat{{\mathcal
W}}^J(n,p)=e^{inp}\frac{\left|D^J(n)+C^J\mbox{diag}\left(z_1^{-1},
\ldots,z_N^{-1}\right)\Delta^J(p)\right|}
{\left|D^J(n)+C^J\mbox{diag}\left(z_1^{-1},
\ldots,z_N^{-1}\right)\right|},
\end{equation}
with
\begin{equation}
\Delta^J(p)\equiv
\mbox{diag}\left(\delta^J(z_1;p),\ldots,\delta^J(z_N;p)\right),
\end{equation}
\begin{equation}
\delta^J(z;p)\equiv
1-\frac{z^{-1}-z}{e^{-ip}-z}=\frac{e^{-ip}-z^{-1}}{e^{-ip}-z}.
\end{equation}
Thus we have
\begin{equation}
\hat{{\mathcal W}}^J(n,p)\sim e^{inp}\prod_{j=1}^N\delta^J(z_j;p),\qquad
n\to -\infty.
\end{equation}

From (\ref{ruijsenaars:WJ2}) and (\ref{ruijsenaars:la-}) we now obtain
\begin{equation}
{\mathcal W}^J(n,p)\sim
e^{inp}\prod_{j=1}^N|z_j|\frac{e^{-ip}-z_j^{-1}}{e^{-ip}-z_j},
\qquad n\to -\infty.
\end{equation}
Substituting (\ref{ruijsenaars:znu}), this yields
(\ref{ruijsenaars:WJ-})--(\ref{ruijsenaars:aJ}),
as announced.

With the above $\eta\to 0$ limits at our disposal, it is
straightforward to calculate the limits of the time-dependent
quantities and time derivatives in Section~2. There is however
one crucial change to be made before doing so: We should
replace $t$ by $it$ so as to obtain the pertinent real-valued
Toda lattice quantities for $\eta \to 0$.

Indeed, doing so in the formula (\ref{ruijsenaars:mut}) that defines the
time-dependence, one obtains
\begin{equation}
\lim_{\eta\to 0} D(ie^{-i\eta}n,it)=\mbox{diag}
\left(1/z_1^{2n+1}\nu_1(t),\ldots,1/z_N^{2n+1}\nu_N(t)\right)\equiv
D^J(n,t),
\end{equation}
\begin{equation}\label{ruijsenaars:nut}
\nu_j(t)\equiv \nu_j\exp\left(-t\left(z_j-z_j^{-1}\right)\right),\qquad
j=1,\ldots,N.
\end{equation}
The time-dependent version of (\ref{ruijsenaars:LJ}) then yields the limit
\begin{equation}
L^J(n,t)_{jk}=C^J_{jk}z_k^{2n}\nu_k(t),\qquad j,k=1,\ldots,N,\qquad
n\in{\mathbb Z},
\end{equation}
and (\ref{ruijsenaars:tauJ}) is generalized to
\begin{equation}
\lim_{\eta\to 0} \tau\left(ie^{-i\eta}n,it\right)=\left|{\bf
1}_N+L^J(n,t)\right|\equiv \tau^J(n,t).
\end{equation}
Then (\ref{ruijsenaars:tauto}) entails Hirota's formula~\cite{ruijsenaars:hiro}
\begin{equation}
\partial_t^2\ln \left(\tau^J(n,t)\right)=\frac{\tau^J(n+1,t)\tau^J(n-1,t)}
{\tau^J(n,t)^2}-1.
\end{equation}
Likewise, (\ref{ruijsenaars:Vad}) and (\ref{ruijsenaars:Vbd}), together
with (\ref{ruijsenaars:an})
and (\ref{ruijsenaars:bn}), yield
\begin{equation}\label{ruijsenaars:ad}
\dot{a}(n,t)=a(n,t)[b(n,t)-b(n+1,t)],
\end{equation}
\begin{equation}\label{ruijsenaars:bd}
\dot{b}(n,t)=2a(n-1,t)^2-2a(n,t)^2.
\end{equation}

Comparing with Flaschka's paper Ref.~\cite{ruijsenaars:flas}, we see that
(\ref{ruijsenaars:ad}), (\ref{ruijsenaars:bd}) coincide with his Eq.~(2.3),
up to
signs. This sign difference arises from our different sign
convention for the time-dependence in (\ref{ruijsenaars:nut}). (Our
convention agrees with Refs.~\cite{ruijsenaars:toda,ruijsenaars:fata}.)
Denoting the
quantities he uses in his Section~3 with a superscript $F$, one
gets (again up to irrelevant conventions)
\begin{equation}
\nu_j=\left(c_j^F\right)^2,\qquad j=1,\ldots,N,
\end{equation}
\begin{equation}
R_j^J(n)=-c_j^Fz_j^{n+1}A_j^{Fn},\qquad j=1,\ldots,N,\qquad
n\in{\mathbb Z},
\end{equation}
\begin{equation}
\lambda^J(n)=K^F(n,n)^{-2}.
\end{equation}
(Compare his system Eq.~(3.3) to (\ref{ruijsenaars:sysJ}) and note his
Eq.~(3.7) to check these correspondences.)

Various formulas in Refs.~\cite{ruijsenaars:toda,ruijsenaars:fata} can also be
obtained as $\eta\to 0$ limits.  In this connection we mention
in particular the Toda/Kac-van Moerbeke account in Section~3.8
of Toda's monograph Ref.~\cite{ruijsenaars:toda}: The pertinent formulas
can be readily derived by taking the $\eta \to 0$ limit of
results that can be found in Section~3 of Part~I.

Furthermore, we point out that van Diejen recently obtained the
analog of the formu\-la~(\ref{ruijsenaars:Wh2}) for the Jacobi wave function
${\mathcal W}^J(n,p)$, cf.~Ref.~\cite{ruijsenaars:vand}. His paper also
contains
further results of interest, and information on recent
literature dealing with Jacobi operators and the
Toda/Kac-van Moerbeke correspondence.

\setcounter{equation}{0}
\section{Parametrization via relativistic Calogero--Moser
systems}

Returning to the general setting of Section~2, we proceed to
detail the connection with the $\tilde{{\rm II}}_{{\rm
rel}}(\tau =\pi/2)$ systems from Ref.~\cite{ruijsenaars:aa2}. The
connection to these finite-dimensional soliton systems hinges
on a suitable reparametrization of the Cauchy matrix (\ref{ruijsenaars:Cr})
and the multipliers in the diagonal matrix $D(r,\mu;x)$
(\ref{ruijsenaars:defD}). To ease the notation we assume from now on that
the numbers $r_1,\ldots,r_N$ are ordered such that
$r_1,\ldots,r_{N_{+}}$ have imaginary parts in $(0,\pi)$ and
$r_{N-N_{-}+1},\ldots,r_N$ in $(-\pi,0)$, with
\begin{equation}
N_{+}\in \{ 0,1,\ldots,N\} ,\qquad N_{-}=N-N_{+}.
\end{equation}
(Since the wave function ${\mathcal W}(r,\mu;x,p)$ and A$\Delta$O
$A(r,\mu)$ are invariant under permutations on the data
$(r,\mu)$, this does not give rise to a loss of generality.)

It now turns out that the numbers $r_1,\ldots,r_{N_{+}}$ and
$r_{N-N_{-}+1},\ldots,r_N$ can be traded for the (complex)
positions $q_1^{+},\ldots,q_{N_{+}}^{+}$ and
$q_1^{-},\ldots,q_{N_{-}}^{-}$ of the particles and
antiparticles, resp., in the $\tilde{{\rm II}}_{{\rm
rel}}$ system, in such a way that the reparametrized Cauchy
matrix $C(r)$ (\ref{ruijsenaars:Cr}) and the Cauchy matrix ${\mathcal C}$
in the $\tilde{{\rm
II}}_{{\rm
rel}}$ Lax matrix are closely related. This relation then
suggests a reparametrization of the multipliers in $D(x)$ such
that the matrix $CD(x)^{-1}$ may be reinterpreted as the  $\tilde{{\rm
II}}_{{\rm
rel}}(\tau =\pi/2)$ Lax matrix evaluated in $x$-dependent
points of the (complexified)  $\tilde{{\rm II}}_{{\rm
rel}}(\tau =\pi/2)$ phase space.

More precisely, the latter identification holds true up to
diagonal similarity transformations. Since we are dealing with
determinants and spectra, such ``gauge transformations'' are
immaterial. In particular, the definition Eq.~(2.70) of the Lax
matrix in Ref.~\cite{ruijsenaars:aa2} amounts to a gauge choice that
facilitates its spectral analysis, but in the present context
another gauge is more convenient.

Specifically, here we work
with the Lax matrix
\begin{equation}\label{ruijsenaars:cL}
{\mathcal L}(q,\theta)\equiv
{\mathcal C}(q^{+},q^{-}){\mathcal D}(q^{+},q^{-},\theta^{+},\theta^{-}).
\end{equation}
The Cauchy matrix ${\mathcal C}$ is defined by
\begin{equation}\label{ruijsenaars:cC1}
{\mathcal C}_{jk}\equiv 1/\cosh [(q_j^{+}-q_k^{+})/2],
\end{equation}
\begin{equation}\label{ruijsenaars:cC2}
{\mathcal C}_{N_{+}+l,N_{+}+m}\equiv 1/\cosh [(q_l^{-}-q_m^{-})/2],
\end{equation}
\begin{equation}\label{ruijsenaars:cC3}
{\mathcal C}_{N_{+}+l,k}\equiv -i/\sinh [(q_l^{-}-q_k^{+})/2],
\end{equation}
\begin{equation}\label{ruijsenaars:cC4}
{\mathcal C}_{j,N_{+}+m}\equiv i/\sinh [(q_j^{+}-q_m^{-})/2].
\end{equation}
Here and from now on, the indices $j$, $k$ take values
$1,\ldots,N_{+}$, whereas the indices $l$, $m$ take values
$1,\ldots,N_{-}$. The diagonal matrix ${\mathcal D}$ is defined by
\begin{equation}\label{ruijsenaars:cD}
{\mathcal D}\equiv\mbox{diag}\left(\exp(\theta_1^{+})V_1^{+},\ldots,
\exp(\theta_{N_{+}}^{+})V_{N_{+}}^{+},
\exp(\theta_1^{-})V_1^{-},\ldots,
\exp(\theta_{N_{-}}^{-})V_{N_{-}}^{-}\right).
\end{equation}
The quantities $\theta_n^{\delta}$ are the generalized momenta
corresponding to the positions $q_n^{\delta}$, and the
``potentials'' $V_n^{\delta}$ are given by
\begin{equation}\label{ruijsenaars:V+}
V_j^{+}\equiv \prod_{1\le k\le N_{+},k\ne j}\left|\coth
[(q_j^{+}-q_k^{+})/2]\right|
\prod_{1\le l\le N_{-}}\left|\tanh [(q_j^{+}-q_l^{-})/2]\right|,
\end{equation}
\begin{equation}\label{ruijsenaars:V-}
V_l^{-}\equiv \prod_{1\le m\le N_{-},m\ne l}\left|\coth
[(q_l^{-}-q_m^{-})/2]\right|
\prod_{1\le j\le N_{+}}\left|\tanh [(q_l^{-}-q_j^{+})/2]\right|.
\end{equation}
(The moduli we are choosing here preclude analyticity in $q$, but
they enable us to steer clear of multi-valuedness issues. Such
issues are important in other contexts, but here they would give
rise to unnecessary complications.)

Comparing (\ref{ruijsenaars:cL})--(\ref{ruijsenaars:V-}) with real $q$'s
and $\theta$'s
to the Lax matrix $L$ given by Eq.~(2.70) in Ref.~\cite{ruijsenaars:aa2},
we see that ${\mathcal L}$ amounts to a diagonal similarity transform, as
announced. More precisely, we should substitute $\tau =\beta\mu
g/2=\pi/2$, $\mu =1$, $\beta =1$, and $x_n^{\delta}\to q_n^{\delta}$,
$p_n^{\delta}\to\theta_n^{\delta}$ in {\it loc.~cit.}, and choose
distinct $q_1^{+},\ldots,q_{N_{-}}^{-}$, cf.~also Eqs.~(1.3)
and (1.4) in Ref.~\cite{ruijsenaars:aa2}. (To avoid confusion, we should
add that the Cauchy matrix ${\mathcal C}$
(\ref{ruijsenaars:cC1})--(\ref{ruijsenaars:cC4}) is
slightly different from the matrix we refer to as Cauchy matrix
in Ref.~\cite{ruijsenaars:aa2}: The latter equals ${\mathcal E}{\mathcal
C}{\mathcal E}$, with
\begin{equation}
{\mathcal
E}\equiv\mbox{diag}\left(\exp(-q_1^{+}/2),\ldots,\exp(-q_{N_{+}}^{+}/2),
\exp(-q_1^{-}/2),\ldots,\exp(-q_{N_{-}}^{-}/2)\right),
\end{equation}
cf.~also Eq.~(B1) in Ref.~\cite{ruijsenaars:aa2}.)

After these preliminaries, we turn to the reparametrizations of
$C(r)$ and the multipliers in $D(r,\mu;x)$. To this end we
first introduce parameters
\begin{equation}\label{ruijsenaars:ka+}
\alpha_j^{+}\equiv -ir_j,
\end{equation}
\begin{equation}\label{ruijsenaars:ka-}
\alpha_l^{-}\equiv -ir_{N_{+}+l}+\pi,
\end{equation}
which are convenient in their own right. Clearly, one has
\begin{equation}
\mbox{Re}\; \alpha_n^{\delta}\in(0,\pi),\qquad n=1,\ldots,N_{\delta},\qquad
\delta =+,-,
\end{equation}
and the Cauchy matrix (\ref{ruijsenaars:Cr}) can be rewritten as
\begin{equation}\label{ruijsenaars:C1}
C_{jk}=-\frac{i}{2}\exp(-i\alpha_j^{+}/2+i\alpha_k^{+}/2)
\frac{1}{\sin[(\alpha_j^{+}+\alpha_k^{+})/2]},
\end{equation}
\begin{equation}\label{ruijsenaars:C2}
C_{N_{+}+l,N_{+}+m}=\frac{i}{2}\exp(-i\alpha_l^{-}/2+i\alpha_m^{-}/2)
\frac{1}{\sin[(\alpha_l^{-}+\alpha_m^{-})/2]},
\end{equation}
\begin{equation}\label{ruijsenaars:C3}
C_{N_{+}+l,k}=-\frac{1}{2}\exp(-i\alpha_l^{-}/2+i\alpha_k^{+}/2)
\frac{1}{\cos[(\alpha_l^{-}+\alpha_k^{+})/2]},
\end{equation}
\begin{equation}\label{ruijsenaars:C4}
C_{j,N_{+}+m}=\frac{1}{2}\exp(-i\alpha_j^{+}/2+i\alpha_m^{-}/2)
\frac{1}{\cos[(\alpha_j^{+}+\alpha_m^{-})/2]}.
\end{equation}
 We also note that the restrictions (\ref{ruijsenaars:req2})
amount to
\begin{equation}\label{ruijsenaars:karest}
\alpha_j^{+}\ne \alpha_k^{+}, \qquad j\ne k,\qquad
\alpha_l^{-}\ne\alpha_m^{-}, \qquad l\ne m,\qquad
\alpha_j^{+}+\alpha_l^{-}\ne \pi.
\end{equation}
(Recall $j,k\in \{ 1,\ldots,N_{+}\}$ and $l,m\in\{
1,\ldots,N_{-}\}$.)

Next, we relate the quantities $\alpha_j^{+}$ and $\alpha_l^{-}$ to
the above positions $q^{+}_j$ and $q_l^{-}$, resp. To this end, we
begin by pointing out that the map $\alpha \mapsto z=\cot (\alpha/2)$
yields an injection of the strip $\mbox{Re}\; \alpha \in(0,\pi)$ onto the
half plane $\mbox{Re}\; z>0$. (Even though this assertion is not immediate,
it is straightforward to check.) Therefore, the map
\begin{equation}\label{ruijsenaars:defF}
F\, :\, \{ \mbox{Re}\; \alpha\in(0,\pi)\} \to \{ \mbox{Im}\;
q\in(-\pi/2,\pi/2)\},\qquad
\alpha\mapsto q=\ln (\cot (\alpha/2)),
\end{equation}
with $F(\pi/2)\equiv 0$, is holomorphic and has a holomorphic
inverse. It is easily seen that the resulting relation
\begin{equation}\label{ruijsenaars:qkap1}
e^{-q}=\tan (\alpha/2),\qquad q=0 \Leftrightarrow \alpha =\pi/2,
\end{equation}
implies
\begin{equation}\label{ruijsenaars:qkap2}
\cosh q=1/\sin\alpha, \qquad \sinh q=\cot\alpha,\qquad \tanh q=\cos\alpha.
\end{equation}

We now set
\begin{equation}\label{ruijsenaars:defq}
q_n^{\delta}\equiv \delta F(\alpha_n^{\delta}),\qquad
n=1,\ldots,N_{\delta},\qquad
\delta =+,-.
\end{equation}
Thus we have
\begin{equation}
\mbox{Im}\; q_n^{\delta}\in (-\pi/2,\pi/2),\qquad
 n=1,\ldots,N_{\delta},\qquad \delta =+,-.
\end{equation}
Moreover, the restrictions (\ref{ruijsenaars:karest}) translate into
\begin{equation}
q_j^{+}\ne q_k^{+},\quad j\ne k,\qquad q_l^{-}\ne q_m^{-},\quad l\ne m,\qquad
q_j^{+}\ne q_l^{-}.
\end{equation}
In words, the positions $q_1^{+},\ldots,
q_{N_{+}}^{+},q_1^{-},\ldots,q_{N_{-}}^{-}$ must be {\em distinct}.

When we combine the changes of variables $r\to
(\alpha^{+},\alpha^{-})\to (q^{+},q^{-})$, we get from the above
\begin{equation}
\exp (r_j)=(i+\sinh q_j^{+})/\cosh q_j^{+},
\end{equation}
\begin{equation}
\exp (r_{N_{+}+l})=(-i+\sinh q_l^{-})/\cosh q_l^{-}.
\end{equation}
But when we would substitute this directly in the Cauchy matrix
(\ref{ruijsenaars:Cr}), we would obtain a matrix whose relation to the Cauchy
matrix ${\mathcal C}$ (\ref{ruijsenaars:cC1})--(\ref{ruijsenaars:cC4}) is
invisible. Instead, we start
from (\ref{ruijsenaars:C1}) and use the relations (\ref{ruijsenaars:qkap1})
and (\ref{ruijsenaars:qkap2}) to calculate
\begin{equation}\arraycolsep=0em
\begin{array}{l}
\displaystyle
\sin (\alpha_j^{+}+\alpha_k^{+})/2   =    \cos(\alpha_j^{+}/2)
\cos(\alpha_k^{+}/2)\left[\tan (\alpha_j^{+}/2)+\tan(\alpha_k^{+}/2)\right]
\vspace{3mm}\\
\displaystyle   \phantom{\sin (\alpha_j^{+}+\alpha_k^{+})/2 }=
\frac{1}{2}\left(
\frac{\sin(\alpha_j^{+})\sin(\alpha_k^{+})}
{\tan(\alpha_j^{+}/2)\tan(\alpha_k^{+}/2)}
\right)^{1/2}
\left[\tan (\alpha_j^{+}/2)+\tan(\alpha_k^{+}/2)\right]
\vspace{3mm}\\
\displaystyle   \phantom{\sin (\alpha_j^{+}+\alpha_k^{+})/2 } =
\frac{\cosh[(q_j^{+}-q_k^{+})/2]}{[\cosh(q_j^{+})\cosh(q_k^{+})]^{1/2}}.
\end{array}
\end{equation}
(Note that the radicands have no zeros or poles in the pertinent
regions, cf.~(\ref{ruijsenaars:defF}). The branch choice is then clear: We need
the positive square root for $q^{+}$ real.)

More generally, in terms of $q^{+}$, $q^{-}$ the Cauchy matrix $C$
(\ref{ruijsenaars:C1})--(\ref{ruijsenaars:C4}) becomes
\begin{equation}\label{ruijsenaars:CcC}
C=S{\mathcal C} S^{-1}D_c,
\end{equation}
where
\begin{equation}
D_c\equiv \frac{i}{2}\,\mbox{diag} \left(-\cosh q_1^{+},\ldots,-\cosh
q_{N_{+}}^{+},
\cosh q_1^{-},\ldots,\cosh q_{N_{-}}^{-}\right),
\end{equation}
\begin{equation}\label{ruijsenaars:defS}
S\equiv \mbox{diag} \left(\exp (-i\alpha_1^{+}/2)[\cosh
q_1^{+}]^{1/2},\ldots,-\exp(-i\alpha_{N_{-}}^{-}/2)[\cosh
q_{N_{-}}^{-}]^{1/2}\right).
\end{equation}
We stick to the parameters $\alpha^{+}$, $\alpha^{-}$ in the exponents,
since this will be convenient shortly.

We now turn to the reparametrization of the multipliers in the
diagonal matrix $D(x)$. In terms of $\alpha^{+}$, $\alpha^{-}$, this
matrix reads
\begin{equation}\label{ruijsenaars:defD2}
D(x)=\mbox{diag} \left(\mu_1(x)\exp (2\alpha_1^{+}x),\ldots,\mu_N(x)\exp
(2\alpha_{N_{-}}^{-}x)\right),
\end{equation}
cf.~(\ref{ruijsenaars:defD})--(\ref{ruijsenaars:defd})
and (\ref{ruijsenaars:ka+})--(\ref{ruijsenaars:ka-}).
Consider now the matrix
\begin{equation}\label{ruijsenaars:defLx}
L(x+i/2)\equiv S^{-1}CD(x)^{-1}S={\mathcal C} D_cD(x)^{-1},
\end{equation}
where we used (\ref{ruijsenaars:CcC}). Comparing it to the Lax matrix
${\mathcal L}$
(\ref{ruijsenaars:cL}), we see that when we rewrite the multipliers as
\begin{equation}\label{ruijsenaars:nu+}
\mu_j(x)=-\frac{i}{2}\exp (i\alpha_j^{+})\cosh
(q_j^{+})\left[V^{+}_j\pi_j^{+}(x+i/2)\right]^{-1},
\end{equation}
\begin{equation}\label{ruijsenaars:nu-}
\mu_{N_{+}+l}(x)=\frac{i}{2}\exp (i\alpha_l^{-})\cosh
(q_l^{-})\left[V_l^{-}\pi_l^{-}(x+i/2)\right]^{-1},
\end{equation}
then we have
\begin{equation}\label{ruijsenaars:LcL}
L(x)={\mathcal L}(q,\theta){\mathcal M}(x),
\end{equation}
with
\begin{equation}\label{ruijsenaars:cM}
{\mathcal M}(x)\equiv\mbox{diag}
\left(\exp(-\theta_1^{+}-2\alpha_1^{+}x)\pi_1^{+}(x),\ldots,\exp
(-\theta_{N_{-}}^{-}-2\alpha_{N_{-}}^{-}x)\pi_{N_{-}}^{-}(x)\right).
\end{equation}

The above formulas (\ref{ruijsenaars:CcC})--(\ref{ruijsenaars:cM}) encode
the announced
relation between the two key matrices $C$ and $D(x)$ and the $\tilde{{\rm
II}}_{{\rm
rel}}(\tau =\pi/2)$ Lax matrix from Ref.~\cite{ruijsenaars:aa2}. Note that the
parametrizations (\ref{ruijsenaars:nu+}) and (\ref{ruijsenaars:nu-}) give
rise to
well-defined $i$-periodic meromorphic multipliers $\pi_n^{\delta}(x)$,
with finite limits
\begin{equation}
\lim_{|{\rm Re}\; x|\to\infty}\pi_n^{\delta}(x)\equiv \pi_n^{\delta},\qquad
n=1,\ldots,N_{\delta},\qquad \delta =+,-.
\end{equation}
Admittedly, at this stage it is not clear that the relation just
established is useful. At first sight, it merely seems a bizarre
coincidence, and indeed the use of the position variables~$q_n^{\delta}$
 would have been quite inconvenient in Part~I and in
Section~2 of the present paper.

As it turns out, however, the relation can be used to great
advantage, not only for studying $N$-soliton solutions (which we do
in Section~6), but also for studying the above A$\Delta$Os from the
viewpoint of quantum mechanics (which we do in Part~III). For both
of these applications, we restrict attention to purely imaginary
$r_1,\ldots,r_N$, and to constant and positive multipliers
$\pi_n^{\delta}$. The associated positions $q_n^{\delta}$ are then real,
and we may and will choose real $\theta_n^{\delta}$ such that
\begin{equation}\label{ruijsenaars:nuth}
\pi_n^{\delta}=\exp \left(\theta_n^{\delta}\right),\qquad
n=1,\ldots,N_{\delta},\qquad \delta
=+,-.
\end{equation}
(Notice that these restrictions amount to the ones at the end of
Section~2.)

With the choices just detailed in effect from now on, we obtain from
each point in the phase space
\begin{equation}\label{ruijsenaars:Omega}\arraycolsep=0em
\begin{array}{l}
\Omega     \equiv   \left\{ (q_1^{+},\ldots,
q_{N_{-}}^{-},\theta^{+}_1,\ldots,\theta_{N_{-}}^{-})\in{\mathbb R}^{2N} \mid
q_{N_{+}}^{+}<\cdots <q_1^{+}, \right.
\vspace{2mm}\\
\phantom{Q\equiv{}}  \left. q_{N_{-}}^{-}<\cdots <q_1^{-},
q_j^{+}\ne q_l^{-},\ j=1,\ldots,N_{+},\ l=1,\ldots,N_{-}\right\}
\end{array}
\end{equation}
of the $\tilde{{\rm II}}_{{\rm rel}}(\tau
=\pi/2)$ system~\cite{ruijsenaars:aa2} an A$\Delta$O $A$ that is formally
self-adjoint, and a real-valued, real-analytic solution to
(\ref{ruijsenaars:nonto}). We now turn to a study of the latter solutions.

\setcounter{equation}{0}
\section{A close-up of the $\pbf{N}$-soliton solutions}

The $\tau$-function (\ref{ruijsenaars:tau}) associated to a point
$(q,\theta)$ in the phase space $\Omega$ (\ref{ruijsenaars:Omega}) can
be rewritten as
\begin{equation}
\tau(x,t)=|{\bf 1}_N+L(x+i/2,t)|=|{\bf 1}_N+L(x,t)U|,
\end{equation}
with
\begin{equation}\label{ruijsenaars:Lxt}\arraycolsep=0em
\begin{array}{l}
L(x,t)\equiv {\mathcal L}\left(q^{+},q^{-},\theta_1^{+}
-2\alpha_1^{+}[x-v(\alpha_1^{+})t],\ldots,
\theta_{N_{-}}^{-}
-2\alpha_{N_{-}}^{-}[x+v(\alpha_{N_{-}}^{-})t]\right),
\vspace{2mm}\\
v(\alpha)\equiv \alpha^{-1}\sin\alpha ,
\end{array}
\end{equation}
\begin{equation}
U\equiv \mbox{diag}
\left(\exp(-i\alpha_1^{+}),\ldots,\exp(-i\alpha_{N_{-}}^{-})\right).
\end{equation}
(Recall (\ref{ruijsenaars:nu+})--(\ref{ruijsenaars:nuth}) and
(\ref{ruijsenaars:mut}) to see this.) The corresponding solution
(\ref{ruijsenaars:Psi}) to (\ref{ruijsenaars:nonto2}) then reads
\begin{equation}\label{ruijsenaars:Nsol}
\Psi(x,t)=i\ln \left(
\frac{|{\bf 1}_N+L(x,t)U^{-1}|}{|{\bf 1}_N+L(x,t)U|}\right) .
\end{equation}

In particular, for $N_{+}=1,N_{-}=0$, one obtains
from (\ref{ruijsenaars:cL})--(\ref{ruijsenaars:cC2})
the right-moving soliton
$f^{+}(q^{+},\theta^{+};x-v(\alpha^{+})t)$, where
\begin{equation}\label{ruijsenaars:1right}
f^{+}(q,\theta;x)\equiv i\ln \left(
\frac{1+e^{i\alpha^{+}}
\exp(\theta -2\alpha^{+}x)}
{1+e^{-i\alpha^{+}}
\exp(\theta -2\alpha^{+}x)}\right)  ,
\end{equation}
\begin{equation}
\alpha^{+}(q)\equiv 2 \,{\rm Arctan}\, (\exp(-q)),
\end{equation}
and for $N_{+}=0,N_{-}=1$, the left-moving soliton
$f^{-}(q^{-},\theta^{-};x+v(\alpha^{-})t)$, where
\begin{equation}\label{ruijsenaars:1left}
f^{-}(q,\theta;x)\equiv i\ln \left(
\frac{1+e^{i\alpha^{-}}
\exp(\theta -2\alpha^{-}x)}
{1+e^{-i\alpha^{-}}
\exp(\theta -2\alpha^{-}x)}\right) ,
\end{equation}
\begin{equation}
\alpha^{-}(q)\equiv 2 \,{\rm Arctan} \, (\exp(q)).
\end{equation}
(We recall that we have already studied these solutions at the
end of Section~2.) We proceed by showing that the general
$N$-soliton solution
$\Psi(x,t)$ (\ref{ruijsenaars:Nsol}) has a long-time asymptotics that
is a linear superposition of $N$ 1-soliton solutions.

To this end we define
\begin{equation}\label{ruijsenaars:Psidel}\arraycolsep=0em
\begin{array}{l}
\displaystyle \Psi^{(\delta)}(x,t)    \equiv
\sum_{j=1}^{N_{+}}f^{+}\left(q^{+}_j,\theta_j^{+};x+\delta
\Delta_j^{+}(q^{+},q^{-})/4\alpha_j^{+}-v(\alpha_j^{+})t\right)
\vspace{3mm}\\
\displaystyle \phantom{\Psi^{(\delta)}(x,t)    \equiv{}}   +
\sum_{l=1}^{N_{-}}f^{-}\left(q^{-}_l,\theta_l^{-};x+\delta
\Delta_l^{-}(q^{+},q^{-})/4\alpha_l^{-}+v(\alpha_l^{-})t\right),
\end{array}
\end{equation}
where $\delta =+,-$ and
\begin{equation}\label{ruijsenaars:Del+}\arraycolsep=0em
\begin{array}{l}
\displaystyle \Delta_j^{+}(q^{+},q^{-})\equiv
\left(\sum_{k<j}-\sum_{k>j}\right)\ln\left(\coth^2[(q_j^{+}-q_k^{+})/2]\right),t)
\vspace{3mm}\\
\displaystyle \phantom{\Delta_j^{+}(q^{+},q^{-})\equiv{}}
- \sum_{l=1}^{N_{-}}\ln\left(\tanh^2[(q_j^{+}-q_l^{-})/2]\right),
\end{array}
\end{equation}
\begin{equation}\label{ruijsenaars:Del-}\arraycolsep=0em
\begin{array}{l}
\displaystyle \Delta_l^{-}(q^{+},q^{-})\equiv
\left(\sum_{m<l}-\sum_{m>l}\right)\ln\left(\coth^2[(q_l^{-}-q_m^{-})/2]\right)
\vspace{3mm}\\
\displaystyle \phantom{\Delta_l^{-}(q^{+},q^{-})\equiv{}}
+ \sum_{j=1}^{N_{+}}\ln\left(\tanh^2[(q_l^{-}-q_j^{+})/2]\right).
\end{array}
\end{equation}
Moreover, we introduce the remainder functions
\begin{equation}
R^{(\delta)}(x,t)\equiv \Psi(x,t)-\Psi^{(\delta)}(x,t),\qquad
\delta =+,-.
\end{equation}
Setting
\begin{equation}
\rho^{(\delta)}(t)\equiv \sup_{x\in{\mathbb R}}\left|R^{(\delta)}(x,t)\right|,
\end{equation}
we conjecture that one has
\begin{equation}
\rho^{(\delta)}(t)=O(\exp(-\delta tr)),\qquad \delta t\to \infty,\qquad \delta
=+,-, \qquad (?)
\end{equation}
with
\begin{equation}\arraycolsep=0em
\begin{array}{l}
r     \equiv   \min\limits_{j\ne k,l\ne
m}\left(2\alpha_j^{+}|v(\alpha_j^{+})-v(\alpha_k^{+})|,
2\alpha_j^{+}(v(\alpha_j^{+})+v(\alpha_l^{-})),\right.
\vspace{2mm}\\
\phantom{r\equiv{}}  \left. 2\alpha_l^{-}|v(\alpha_l^{-})-v(\alpha_m^{-})|,
2\alpha_l^{-}(v(\alpha_l^{-})+v(\alpha_j^{+}))\right).
\end{array}
\end{equation}
In Section~7 of Ref.~\cite{ruijsenaars:aa2} we proved the analog of
this conjecture for the particle-like solutions to the
sine-Gordon, modified KdV and KdV equations.
Unfortunately, the strategy we followed in the latter
cases does not apply here.

On the other hand, the structure of the solution
$\Psi(x,t)$ (\ref{ruijsenaars:Nsol}) makes it possible to supply a
quite simple and direct proof of a weaker convergence
result, expressed in the following proposition.

\begin{prop}
Fixing $x_0,s_0\in{\mathbb R}$, one has
\begin{equation}\label{ruijsenaars:plim}
\lim_{\delta t\to \infty}\exp\left(-iR^{(\delta)}(x_0+s_0t,t)\right)=1,\qquad
\delta =+,-.
\end{equation}
\end{prop}

In order to appreciate this proposition and to prepare for
its proof, we begin by pointing out that the function
\begin{equation}
F_t(x_0,s_0)\equiv
f^{+}\left(q,\theta;x_0+s_0t-v(\alpha^{+})t\right)
\end{equation}
has the following {\em discontinuous} limiting behavior:
\begin{equation}\label{ruijsenaars:Ftlim}
\lim_{t\to\infty}F_t(x_0,s_0)=
\left\{
\begin{array}{ll}
0,   &   s_0>v(\alpha^{+}),
\vspace{1mm}\\
-2\alpha^{+},  &  s_0<v(\alpha^{+}),
\vspace{1mm}\\
f^{+}(q,\theta;x_0),   &   s_0=v(\alpha^{+}).
\end{array}
\right.
\end{equation}
(Indeed, this is clear from (\ref{ruijsenaars:1right}).) From this the
$t\to \pm\infty$ limits of $\Psi^{(\pm)}(x_0+s_0t,t)$
(cf.~(\ref{ruijsenaars:Psidel})) are quite easily calculated. Denoting
the limits by $F^{(\pm)}(x_0,s_0)$, Proposition~6.1 is
equivalent to
\begin{equation}\label{ruijsenaars:kdiff}
\lim_{\delta
t\to\infty}\Psi(x_0+s_0t,t)=F^{(\delta)}(x_0,s_0)+2\pi k,\qquad
\delta =+,-,\qquad k\in{\mathbb Z}.
\end{equation}
A priori, the integer $k$ depends on $\delta$, $x_0$ and $s_0$,
though it is undoubtedly true that one has $k=0$ (recall
the branch choice (\ref{ruijsenaars:branch})). But we cannot
rigorously deduce this, since (\ref{ruijsenaars:plim}) is a pointwise
limit.

Next, we observe that the $t$-dependence of
$\Psi(x_0+s_0t,t)$ is carried by factors
$\exp(t\lambda_i(s_0))$ in the Lax matrix, with
\begin{equation}\label{ruijsenaars:lami}
\lambda_i(s)\equiv \left\{
\begin{array}{ll}
-2\alpha_i^{+}(s-v(\alpha_i^{+})),  &  i=1,\ldots,N_{+},
\vspace{1mm}\\
-2\alpha_{i-N_{+}}^{-}(s+v(\alpha_{i-N_{+}}^{-})),  &
i=N_{+}+1,\ldots,N,
\end{array}
\right.
\end{equation}
cf.~(\ref{ruijsenaars:Lxt}). Furthermore, all principal minors of the
matrices $LU^{\pm 1}$ occurring in (\ref{ruijsenaars:Nsol}) can be
readily calculated via Cauchy's identity, yielding non-zero
numbers.

We now turn to a lemma in which the state of affairs just
sketched is studied in a somewhat more general setting. For
$M$ an $N\times N$ matrix, denote by $M_j$ the $j\times j$
matrix obtained from $M$ by deleting the rows and columns
$j+1,\ldots,N$. Now we define a set ${\mathcal M}$ of matrices by
\begin{equation}\label{ruijsenaars:mset}
{\mathcal M} \equiv \{ M\in M_N({\mathbb C})\mid |M_j|\ne 0,\ j=1,\ldots,N\} .
\end{equation}

\begin{lem}
Let $M^{+},M^{-}\in{\mathcal M}$ and let
\begin{equation}
Q(t)\equiv \left|{\bf 1}_N+M^{+}e^{tD}\right|/
\left|{\bf 1}_N+M^{-}e^{tD}\right|,
\end{equation}
where
\begin{equation}\label{ruijsenaars:Dd}
D\equiv \mbox{\rm diag}\, (d_1,\ldots,d_N),\qquad d_1,\ldots,d_N\in{\mathbb R}.
\end{equation}
Assuming
\begin{equation}\label{ruijsenaars:A1}
d_1,\ldots,d_n\in (0,\infty),\qquad
d_{n+1},\ldots,d_N\in(-\infty,0),\qquad n\in \{ 0,\ldots,N\},
\end{equation}
one has
\begin{equation}\label{ruijsenaars:lim1}
\lim_{t\to\infty}Q(t)=|M_n^{+}|/|M_n^{-}|,
\end{equation}
with $|M_0|\equiv 1$. Assuming next
\begin{equation}\label{ruijsenaars:A2}\arraycolsep=0em
\begin{array}{l}
d_1,\ldots,d_n\in (0,\infty),\qquad
d_{n+1}=0,
\vspace{1mm}\\
 d_{n+2},\ldots,d_N\in(-\infty,0),\qquad n\in \{ 0,\ldots,N-1 \},
\end{array}
\end{equation}
one has
\begin{equation}\label{ruijsenaars:lim2}
\lim_{t\to\infty}Q(t)=(|M_n^{+}|+|M_{n+1}^{+}|)|/(|M_n^{-}|
+|M_{n+1}^{-}|).
\end{equation}
\end{lem}

\noindent
{\bf Proof of Lemma~6.2.} We may write
\begin{equation}
Q(t)=\eta^{+}(t)/\eta^{-}(t),
\end{equation}
where
\begin{equation}\arraycolsep=0em
\begin{array}{l}
\eta^{\delta}(t)\equiv
\left|\mbox{diag}\left(e^{-td_1},\ldots,e^{-td_n},1,\ldots,1\right)\right.
\vspace{2mm}\\
\phantom{\eta^{\delta}(t)\equiv{}}
\left.
+M^{\delta}\mbox{diag}\left(1,\ldots,1,e^{td_{n+1}},\ldots,e^{td_N}\right)\right
|,
\qquad \delta =+,-.
\end{array}
\end{equation}
Now the assumption (\ref{ruijsenaars:A1}) entails
\begin{equation}\label{ruijsenaars:mlim}
\lim_{t\to\infty}\eta^{\delta}(t)=\left|
\begin{array}{clclll}
M_{11}^{\delta}  &  \cdots  &  M_{1n}^{\delta}  &  0
&  \cdots  &  0  \\
\vdots  &  &  \vdots  &  \vdots  &  &  \vdots \\
M_{n1}^{\delta}  &  \cdots  &  M_{nn}^{\delta}  &  0
&  \cdots  &  0  \\
M_{n+1,1}^{\delta}  &  \cdots  &  M_{n+1,n}^{\delta}  &  1
&  \cdots  &  0  \\
\vdots  &  &  \vdots  &  \vdots  & \ddots &  \vdots \\
M_{N1}^{\delta}  &  \cdots  &  M_{Nn}^{\delta}  &  0
&  \cdots  &  1
\end{array} \right|
=\left|M_n^{\delta}\right|,
\end{equation}
whence (\ref{ruijsenaars:lim1}) is clear. The assumption (\ref{ruijsenaars:A2})
gives rise to the same limit matrix as in
(\ref{ruijsenaars:mlim}), except that one should add
$\left(M_{1,n+1}^{\delta},\ldots,M_{N,n+1}^{\delta}\right)^t$ to the
$(n+1)$th column. A~moment's thought then yields
\begin{equation}
\lim_{t\to\infty}\eta^{\delta}(t)=\left|M_n^{\delta}\right|+\left|M_{n+1}^{\delta}\right|,
\end{equation}
and so (\ref{ruijsenaars:lim2}) results.\hfill \rule{3mm}{3mm}

\medskip

\noindent
{\bf Proof of Proposition~6.1.} We only show
\begin{equation}\label{ruijsenaars:QF}
\lim_{t\to\infty}\frac{\left|{\bf 1}_N+L(x_0+s_0t,t)U^{-1}\right|}
{\left|{\bf 1}_N+L(x_0+s_0t,t)U\right|}=\exp\left(-iF^{(+)}(x_0,s_0)\right),
\end{equation}
the proof for $t\to -\infty$ being similar. Let us first
note that the ordering of the positions entails
\begin{equation}\label{ruijsenaars:veloc}
-v(\alpha_{N_{-}}^{-})<\cdots
<-v(\alpha_1^{-})<v(\alpha_{N_{+}}^{+})<
\cdots
<v(\alpha_1^{+}).
\end{equation}
For $s_0>v(\alpha_1^{+})$ we therefore have
$\lambda_i(s_0)<0$, $i=1,\ldots,N$, cf.~(\ref{ruijsenaars:lami}). Thus
we can invoke Lemma~6.2 with
\begin{equation}\label{ruijsenaars:ML}
M^{\pm}=L(x_0,0)U^{\mp 1},
\end{equation}
and $n=0$ in (\ref{ruijsenaars:A1}) to obtain limit 1 on the lhs of
(\ref{ruijsenaars:QF}).
Since we have
\begin{equation}
F^{(+)}(x_0,s_0)=0,\qquad s_0>v(\alpha_1^{+}),
\end{equation}
we obtain (\ref{ruijsenaars:QF}) for $s_0>v(\alpha_1^{+})$.

For $s_0\in (v(\alpha_2^{+}),v(\alpha_1^{+}))$ we have
$\lambda_1(s_0)>0$, with all other $\lambda_i(s_0)$
still negative. Thus Lemma~6.2 applies with (\ref{ruijsenaars:ML})
in effect, and with $n=1$ in (\ref{ruijsenaars:A1}). Then the lhs of
(\ref{ruijsenaars:QF}) yields $\exp(2i\alpha_1^{+})$, which equals
the rhs (recall (\ref{ruijsenaars:Ftlim})). Likewise, it follows
that (\ref{ruijsenaars:QF}) holds true for $s_0$ not equal to the
velocities (\ref{ruijsenaars:veloc}).

Next, we choose $s_0=v(\alpha_1^{+})$. Then we have
$\lambda_1(s_0)=0$, with all other $\lambda_i(s_0)$
negative. Thus we can exploit Lemma~6.2 with $n=0$ in
(\ref{ruijsenaars:A2}), which yields a limit
\begin{equation}
\frac{1+(L(x_0,0)U^{-1})_{11}}{1+(L(x_0,0)U)_{11}}=
\frac{1+\exp(\theta_1^{+}-2\alpha_1^{+}x_0)V_1^{+}
\exp(i\alpha_1^{+})}
{1+\exp(\theta_1^{+}-2\alpha_1^{+}x_0)V_1^{+}
\exp(-i\alpha_1^{+})}.
\end{equation}
This agrees with the rhs of (\ref{ruijsenaars:QF}), cf.~(\ref{ruijsenaars:V+}),
(\ref{ruijsenaars:Psidel}) and (\ref{ruijsenaars:Del+}).

Choosing now $s_0=v(\alpha_2^{+})$, Lemma~6.2 with $n=1$
in (\ref{ruijsenaars:A2}) yields the limit
\begin{equation}
\exp(2i\alpha_1^{+})\cdot \frac{1+m_2\exp(i\alpha_2^{+})}
{1+m_2\exp(-i\alpha_2^{+})},
\end{equation}
where $m_2$ denotes the quotient of the principal minor
of $L(x_0,0)$ with respect to indices~1,~2 and its
$11$-element. Cauchy's identity yields
\begin{equation}
m_2=\exp(\theta_2^{+}-2\alpha_2^{+}x_0)V_2^{+}
\tanh^2[(q_1^{+}-q_2^{+})/2],
\end{equation}
and so the limit agrees with the rhs, viz.,
\begin{equation}
\exp\left[2i\alpha_1^{+}-if^{+}(q_2^{+},\theta_2^{+};x_0+\Delta_2^{+}(q^{+},q^{-
})
/4\alpha_2^{+})\right],
\end{equation}
cf.~(\ref{ruijsenaars:Psidel}), (\ref{ruijsenaars:Del+}), and
(\ref{ruijsenaars:Ftlim}).

Proceeding in the same way for $s_0$ equal to the
remaining velocities, it is readily checked that
(\ref{ruijsenaars:QF}) holds true as well. Therefore,
Proposition~6.1 now follows.
\hfill \rule{3mm}{3mm}

\medskip

Finally, we briefly consider the above $N$-soliton
solutions from the perspective of Section~7 in
Ref.~\cite{ruijsenaars:aa2}, especially as concerns the issue of
soliton space-time trajectories. Inspec\-ting the
derivatives (\ref{ruijsenaars:der+}) and (\ref{ruijsenaars:der-}), we see that
there is an obvious choice of space-time trajectories
for the 1-soliton solutions, namely,
\begin{equation}\label{ruijsenaars:tr+}
x^{+}(t)=\theta^{+}/2\alpha^{+}+v(\alpha^{+})t,
\end{equation}
\begin{equation}\label{ruijsenaars:tr-}
x^{-}(t)=\theta^{-}/2\alpha^{-}-v(\alpha^{-})t.
\end{equation}

Let us now choose $N_{+}=N$, $N_{-}=0$ until further
notice. Then we are dealing with the self-dual
${\rm II}_{{\rm
rel}}(\tau =\pi/2)$ system. For  the soliton solutions
arising in Section~7 of Ref.~\cite{ruijsenaars:aa2} this
specialization amounts to the pure soliton case studied
in Subsection~7A. In order to compare the present
setting to {\it loc.~cit.}, we should take $\beta
=1$, $\tau =\pi /2$ in {\it loc.~cit.}, and substitute
$q$, $\theta \to \hat{\theta}$, $\hat{q}$. Then the above
Lax matrix (\ref{ruijsenaars:cL}) turns into a diagonal similarity
transform of the Lax matrix (7.4) in {\it loc.~cit.}

Substituting next in {\it loc.~cit.}
\begin{equation}
y\to x, \qquad \sigma_j=2\alpha_j^{+},\qquad
v_j=v(\alpha_j^{+}),
\end{equation}
one gets agreement with Eq.~(7.7). Since $L(x,t)$
(\ref{ruijsenaars:Lxt}) amounts to a diagonal similarity transform
of the matrix $\tilde{A}(t,x)$ (7.6), we may apply
Theorem~7.1 to the case at hand. Thus we obtain
non-intersecting soliton space-time trajectories with
long-time asymptotics
\begin{equation}\label{ruijsenaars:traj}
x^{+}_{\stackrel{\scriptstyle{j}}{\scriptstyle{N-j+1}}}(t)=\frac{1}
{2\alpha^{+}_j}\left( \theta_j^{+}\mp
\frac{1}{2}\Delta_j^{+}(q^{+})\right)
+v(\alpha_j^{+})t+O(\exp(\mp tr_j)),\qquad t\to \pm
\infty,
\end{equation}
where
\begin{equation}
\Delta_j^{+}(q^{+})\equiv
\left(\sum_{k<j}-\sum_{k>j}\right)\ln\left(\coth^2[(q_j^{+}-q_k^{+})/2]\right),\
qquad
 j=1,\ldots,N,
\end{equation}
\begin{equation}
r_j\equiv \min_{k\ne j}2\alpha_k^{+}|v(\alpha_k^{+})-v(\alpha_j^{+})|,\qquad
j=1,\ldots,N.
\end{equation}
(See Ref.~\cite{ruijsenaars:ober} for a picture of sine-Gordon
soliton space-time trajectories.)

To be sure, without prior knowledge it would  not at all
be obvious that the terminology used here is appropriate
for the solution (\ref{ruijsenaars:Nsol}). Indeed, since $L(x,t)$ is
multiplied by the $q^{+}$-dependent phase matrices $U$
and $U^{*}$, we have no analog of the ``1-particle
superposition'' formulas Eqs.~(7.10)--(7.14) in {\it
loc.~cit.}. More importantly, at face value the
long-time asymptotics of the spectrum of $L(x,t)$ seems
to have no bearing on the long-time asymptotics of the
solution $\Psi(x,t)$ (\ref{ruijsenaars:Nsol}).

Even so, we need only invoke Proposition~6.1 for the
case $N_{+}=N$, $N_{-}=0$ at hand to deduce that the
asymptotic space-time trajectories in (\ref{ruijsenaars:traj})
coincide with the 1-soliton trajectories in
(\ref{ruijsenaars:Psidel}), cf.~also (\ref{ruijsenaars:tr+}). Accordingly, the
trajectories in (\ref{ruijsenaars:traj}) do encode the physical
characteristics of the right-moving $N$-soliton
solutions.

Likewise, for the case $N_{+}=0$, $N_{-}=N$ we can use
Theorem~7.1 in {\it loc.~cit.} to obtain non-intersecting space-time
trajectories that coincide with the obvious ``locations''
of the $N$ left-moving solitons for asymptotic times.

Whenever $N_{+}N_{-}>0$, however, Theorem~7.1 can no
longer be used to define space-time trajectories.
This is because $L(x,t)$ need not have positive and
simple spectrum when particles and antiparticles are
present. Indeed, as
$(q^{+},q^{-},\theta^{+},\theta^{-})$ varies over
$\Omega$ (\ref{ruijsenaars:Omega}), the eigenvalues of $L(0,0)$ range
over all of the right half plane, cf.~{\it loc.~cit.},
Subsection~2C.

As an illuminating example, consider the case
$N_{+},N_{-}=1$ with $q\equiv (q^{+}-q^{-})/2$:
\begin{equation}\label{ruijsenaars:examp}
L(x,t)=\left(
\begin{array}{cc}
\tanh |q|  &  i/\cosh q  \\
i/\cosh q &  \tanh |q|
\end{array}
\right)\cdot
\left(
\begin{array}{cc}
\exp(\theta^{+}(x,t))  &  0 \\
0 &  \exp(\theta^{-}(x,t))
\end{array}
\right),
\end{equation}
\begin{equation}
\theta^{\delta}(x,t)\equiv \theta^{\delta}
-2\alpha^{\delta}\left(x-\delta v(\alpha^{\delta})t\right),\qquad \delta =+,-.
\end{equation}
Setting
\begin{equation}\label{ruijsenaars:sigm}
\sigma(x,t)\equiv
\tanh^2([\theta^{+}(x,t)-\theta^{-}(x,t)]/2)-1/\cosh^2q,
\end{equation}
it is routine to verify that
for $\alpha^{+}\ne \alpha^{-}$ the matrix $L(x,t)$ has
two distinct positive eigenvalues  when
$\sigma(x,t)>0$ and a complex-conjugate pair of
eigenvalues in the right half plane when $\sigma(x,t)<0$,
whereas
$L(x,t)$ is not diagonalizable when $\sigma(x,t)$
vanishes.
For an {\em arbitrary} fixed $t_0$, one therefore finds
that $L(x,t_0)$ has two distinct positive eigenvalues
for $x>x_0^{+}$ and $x<x_0^{-}$, and non-real spectrum
for $x\in (x_0^{-},x_0^{+})$. Moreover, in the
non-generic case $\alpha^{+}=\alpha^{-}$ the spectral
character only depends on $t_0$. (As a bonus, these
observations show that the Hamiltonians on $\Omega$ that
generate the $x$- and $t$-flows generically do not leave
the spectral decomposition of $\Omega$ invariant.)

In spite of this quite different state of affairs for
$N_{+}N_{-}>0$, we believe that a more refined spectral
analysis of $L(x,t)$ should yield $N$ non-intersecting
space-time trajectories for $t>T^{(+)}$ and
$t<T^{(-)}$, where $T^{(\pm)}$ depend only on
$(q,\theta)\in \Omega$ (\ref{ruijsenaars:Omega}). The asymptotics
of these trajectories should read
\begin{equation}\label{ruijsenaars:x+}
x^{+}_{\stackrel{\scriptstyle{j}}
{\scriptstyle{N_{+}-j+1}}}(t)\sim \frac{1}
{2\alpha^{+}_j}\left( \theta_j^{+}\mp
\frac{1}{2}\Delta_j^{+}(q^{+},q^{-})\right)
+v(\alpha_j^{+})t,\qquad t\to \pm
\infty,
\end{equation}
\begin{equation}\label{ruijsenaars:x-}
x^{-}_{\stackrel{\scriptstyle{l}}
{\scriptstyle{N_{-}-l+1}}}(t)\sim \frac{1}
{2\alpha^{-}_l}\left( \theta_l^{-}\mp
\frac{1}{2}\Delta_l^{-}(q^{+},q^{-})\right)
-v(\alpha_l^{-})t,\qquad t\to \pm \infty,
\end{equation}
(with $\Delta_j^{+}$ and $\Delta_l^{-}$ given by (\ref{ruijsenaars:Del+}) and
(\ref{ruijsenaars:Del-})), so that the trajectories coincide
asymptotically with the 1-soliton trajectories following
from (\ref{ruijsenaars:Psidel}).

To explain what is involved here, it is illuminating to
reconsider the example (\ref{ruijsenaars:examp})--(\ref{ruijsenaars:sigm}) with
$q^{+}=-q^{-}$ (so that $\alpha^{+}=\alpha^{-}$). Then one
readily verifies the following features. First, there exist
unique $T^{(+)}$, $T^{(-)}$ such that $L(x,t)$ has non-real
spectrum for $t\in (T^{(-)},T^{(+)})$ and distinct positive
eigenvalues for $t>T^{(+)}$ and $t<T^{(-)}$. Second, for
$t>T^{(+)}$ there exist unique $x^{+}(t)$, $x^{-}(t)$ with
$x^{+}(t)>x^{-}(t)$ such that $L(x^{\delta}(t),t)$, $\delta =+,-,$ has
non-degenerate eigenvalue~1. Likewise, for $t<T^{(-)}$ there
exist unique $x^{+}(t)$, $x^{-}(t)$ with
$x^{+}(t)<x^{-}(t)$ such that $L(x^{\delta}(t),t)$, $\delta =+,-,$ has
non-degenerate eigenvalue~1. Finally, the asymptotics of the
trajectories $x^{\delta}(t)$, $\delta =+,-,$ is given by (\ref{ruijsenaars:x+})
and (\ref{ruijsenaars:x-}). (Physically speaking, the particle and
antiparticle associated with this special two-soliton solution
form a virtual bound state/resonance for $t\in
[T^{(-)},T^{(+)}]$.)

Of course, this example may be viewed as too special to
yield convincing evidence. Indeed, a far more important
reason why it is plausible that trajectories with the
above-mentioned properties exist is the following lemma.

\begin{lem}
Let $M\in {\mathcal M} $ (\ref{ruijsenaars:mset}) and let
\begin{equation}
E(t)\equiv Me^{tD},
\end{equation}
where $D$ is given by (\ref{ruijsenaars:Dd}) and (\ref{ruijsenaars:A2}).
Then there
exists $T\in {\mathbb R}$ such that for all $t\ge T$ the matrix
$E(t)$ has a non-degenerate eigenvalue $e_n(t)$ obeying
\begin{equation}
e_n(t)=m_n(1+\rho_n(t)), \qquad m_n\equiv |M_n|/M_{n-1}|,
\end{equation}
\begin{equation}
|\rho_n(t)|\le C\exp(-tr_n),\qquad r_n\equiv \min
(d_1,\ldots,d_n,-d_{n+2},\ldots,-d_N).
\end{equation}
\end{lem}

\noindent
{\bf Proof}. This lemma can be obtained as a corollary of
the proof of Theorem~A2 in Ref.~\cite{ruijsenaars:aa1}. The latter
theorem concerns the case where $d_1,\ldots,d_N$ are
distinct, which is not implied by our assumption (\ref{ruijsenaars:A2}).
But when one follows the arguments in its proof, one easily
sees that they entail the assertion of the lemma. \hfill \rule{3mm}{3mm}

\medskip

We now explain the bearing of this lemma on our trajectory
conjecture. To this end we consider (for example) the matrix
\begin{equation}
\label{ruijsenaars:Lj}
L(x_0+v(\alpha_j^{+})t,t)=L(x_0,0)\,\mbox{diag} \left(\exp
(\lambda_1(v(\alpha_j^{+}))t),\ldots,
\exp(\lambda_N(v(\alpha_j^{+}))t\right),
\end{equation}
where we used (\ref{ruijsenaars:Lxt}) and (\ref{ruijsenaars:lami}). Taking
$M=L(x_0,0)$ and
$n=j$ in Lemma~6.3, we see that the assumptions are
satisfied. Since the linear functions $\lambda_i(s)$
(\ref{ruijsenaars:lami}) intersect for $s$ varying over a finite set
$\cal I$, the numbers
$d_1,\ldots,d_N$ in the lemma are not distinct whenever
$v(\alpha_j^{+})\in {\cal I}$. This is the reason why
Theorem~A2  in Ref.~\cite{ruijsenaars:aa1} does not apply in general.
(For $s$-values not in $\cal I$, however, Theorem~A2 does
apply to $L(x_0+st,t)$, entailing simple and positive
spectrum for $|t|$ sufficiently large.)

As a consequence, we deduce that for $t>T_j$ the matrix
(\ref{ruijsenaars:Lj}) has a non-degenerate eigenvalue $e_j(x_0,t)$. The
trajectory $x^{+}_j(t)$ should now be defined so that
\begin{equation}
e_j(x_j^{+}(t)-v(\alpha_j^{+})t,t)=1.
\end{equation}
One readily checks that this would yield the desired
asymptotics for $t\to\infty$, but we cannot rigorously prove
that there exists a unique $x_0\in{\mathbb R}$ such that
$e_j(x_0,t)=1$. (For one thing, the choice of $T_j$ {\em
depends} on the $x_0$ we fix.)

In any event, by now it will be clear what spectral
feature our trajectory conjecture amounts to: For large times
$t$, the matrix $L(x,t)$ should have a non-degenerate
eigenvalue~1 for $N$ and only $N$ positions
$x_{N_{-}}^{-}(t)<\cdots <x_1^{-}(t)<x_{N_{+}}^{+}(t)<\cdots
<x_1^{+}(t)$ --- the soliton space-time trajectories. Though
the above arguments do not constitute a complete proof, they
provide considerable evidence. To conclude, it should be
emphasized that for $N_{+}N_{-}>0$ one should not expect
global trajectories. Indeed, as we have already established
explicitly for the above example (\ref{ruijsenaars:examp}) with
$q^{-}=-q^{+}$, it is likely that when left- and
right-moving trajectories collide, the corresponding
eigenvalue pair becomes non-real.

\label{ruijsenaarsII-lastpage}

\newpage

\end{document}